\pdfoutput=1 
\documentclass[aps,prx,superscriptaddress,twocolumn, secnumarabic, amssymb,nobibnotes, floatfix, 10pt, pdftex]{revtex4-2}

\usepackage[T1]{fontenc}
\usepackage[nolist]{acronym}
\usepackage{microtype}
\usepackage{graphicx} % bilder
\usepackage{epstopdf}
\usepackage{grffile} % bilder
\usepackage{lipsum} % help wrapfigure
\usepackage[english]{babel}%rechtschreibung
\usepackage{amsmath}
\usepackage{cancel}
\usepackage{bm}
\usepackage{siunitx}
\usepackage{color}
\usepackage[percent]{overpic}
\usepackage{parskip}
\usepackage{blindtext}
\DeclareSIUnit{\ang}{\mbox{\normalfont\AA}}

\graphicspath{{./figs/}}

\begin{document}

\title{  Proton-transfer spectroscopy beyond the normal-mode scenario}

\author{Florian N. Br\"unig}
\affiliation{Freie Universität Berlin, Department of Physics, 14195 Berlin, Germany}

\author{Paul Hillmann}
\affiliation{Freie Universität Berlin, Department of Physics, 14195 Berlin, Germany}

\author{Won Kyu Kim}
\affiliation{Korea Institute for Advanced Study, School of Computational Sciences, Seoul 02455, Republic of Korea}

\author{Jan O. Daldrop}
\affiliation{Freie Universität Berlin, Department of Physics, 14195 Berlin, Germany}

\author{Roland R. Netz}
\email[]{rnetz@physik.fu-berlin.de}
\affiliation{Freie Universität Berlin, Department of Physics, 14195 Berlin, Germany}

\date{\today}

\begin{abstract}
A stochastic theory is developed to predict the spectral signature of proton transfer processes
and applied to infrared  spectra computed from ab initio molecular-dynamics simulations of a single
 H$_5$O$_2{}^{+}$ cation. By constraining the oxygen atoms to a fixed distance, this system serves as a tunable model for general
 proton-transfer processes with variable barrier height. Three  spectral contributions at distinct frequencies
 are identified and analytically predicted:
 the quasi-harmonic motion around the most probable configuration, amenable to  normal-mode analysis,
 the  contribution due to transfer paths   when the proton moves over  the barrier
 and a shoulder for  low frequencies  stemming from the stochastic transfer-waiting-time distribution;
 the latter two contributions are not captured by normal-mode analysis but exclusively
 report on the proton-transfer kinetics.
 In accordance with reaction kinetic theory, the transfer-waiting-contribution frequency
 depends inverse exponentially on the barrier height, whereas the transfer-path-contribution frequency is rather insensitive to the barrier height.
\end{abstract}

\maketitle

\section{Introduction}

\begin{figure*}
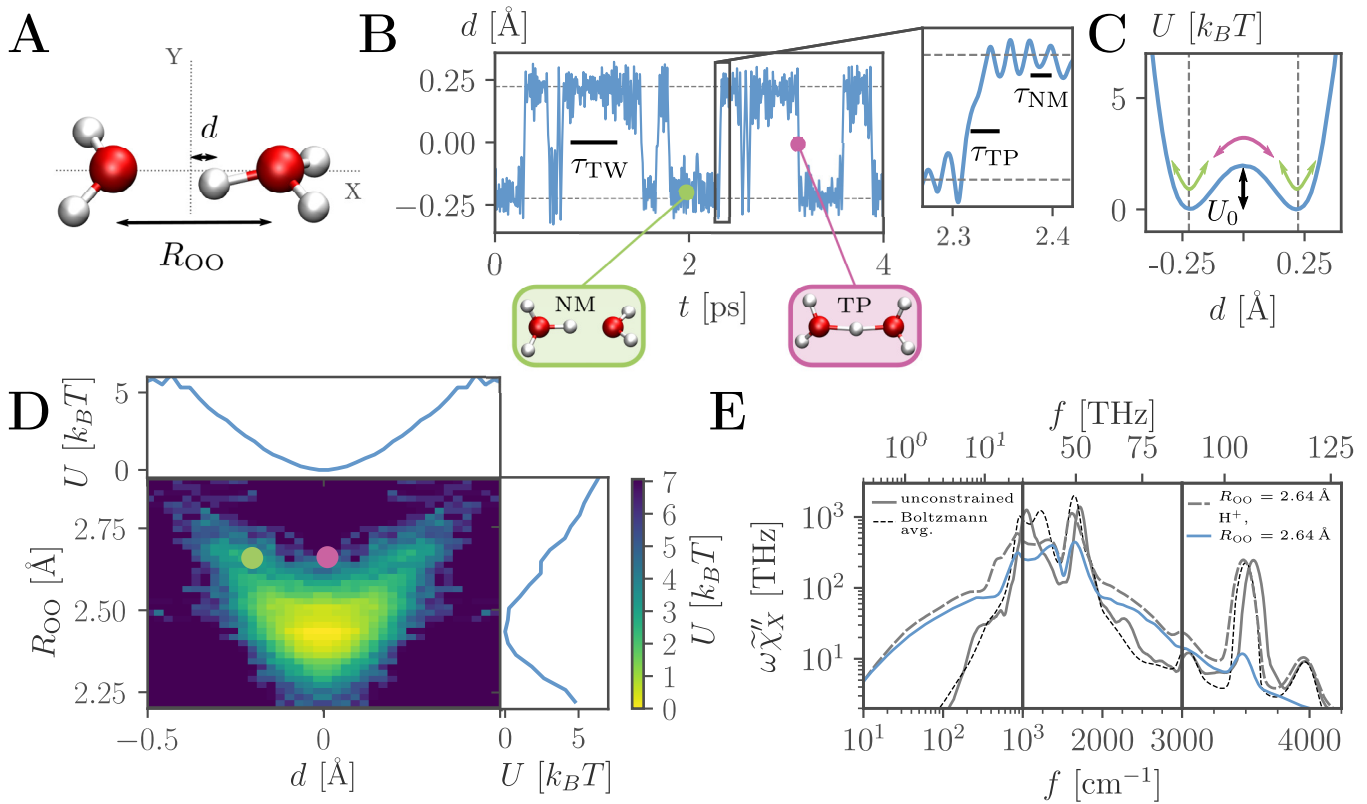

\centering
\begin{overpic}[width=\textwidth]{{/../figs/zundel_intro8}.eps}
\put(0,58){\huge \bf A}
\put(26,58){\huge \bf B}
\put(80,58){\huge \bf C}
\put(0,30){\huge \bf D}
\put(52,30){\huge \bf E}
\end{overpic}
\caption{Ab initio molecular-dynamics (AIMD) simulations of the H$_5$O$_2{}^{+}$  cation. 
\textbf{A} The oxygen-oxygen separation
$R_{\mathrm{OO}}$ and the proton distance from the oxygen midpoint   along the $x$ axis, named $d$, describe  the excess-proton dynamics. 
\textbf{B} The excess proton  trajectory for fixed $R_{\mathrm{OO}}=\SI{2.64}{\ang}$ visualizes the transfer-waiting time  $\tau_{\mathrm{TW}}$ as well as the normal-mode time  $\tau_{\mathrm{NM}}$ and the  transfer-path (TP) time  $\tau_{\mathrm{TP}}$ (see inset).
Selected snapshots show structures at the     free-energy minimum and at the  barrier top.
\textbf{C} Free-energy profile  for fixed $R_{\mathrm{OO}}=\SI{2.64}{\ang}$, extracted from  constrained simulations.
\textbf{D} 2D free-energy landscape  in terms of $R_{\mathrm{OO}}$ and $d$ from unconstrained simulations.
\textbf{E} Absorption  spectra along the $x$  axis where $\omega=2\pi f$.
The grey  solid line shows the total (i.e. nuclear + electronic)
 spectrum of the unconstrained system, compared with the Boltzmann average of
constrained systems (black broken line).
The grey broken line shows the total spectrum for constrained $R_{\mathrm{OO}}=\SI{2.64}{\ang}$,
compared  to the spectrum of only the excess proton (blue line, multiplied  by a factor of 2). Note the change of scales
at $f=1000\,$cm$^{-1}$ and $f=3000\,$cm$^{-1}$.
}
\label{zundel_intro}
\end{figure*}

The transfer dynamics of excess protons in the aqueous environment is central to many biochemical processes \cite{Marx2006},
but despite substantial work,  even for acidic water a complete  kinetic model that would describe  all spectral features encompassing the low THz and \ac{IR}  regimes  remains elusive.
Typically, the  discussion is based  on two  idealized  proton-transfer intermediates,
namely the H$_5$O$_2{}^{+}$  Zundel cation, where two water molecules symmetrically
point their oxygens to the excess proton \cite{Zundel1968},
 and the Eigen cation, where   hydronium H$_3$O$^+$ is formed and solvated by three water molecules \cite{Wicke1954}.
 Accordingly, proton diffusion in water is portrayed as a stochastic succession of these two states, where
  the excess proton  switches during  diffusion:
 It is  a defect that diffuses, rather than a specific proton,
 which explains the  high proton mobility in water \cite{Tuckerman1995, Berkelbach2009}.

An intensely debated question concerns the relative stability  and abundance of the Eigen and Zundel forms in acidic  water \cite{Asthagiri2005,Daly2017,Dahms2017, Carpenter2018, Calio2021}.
Several experimental  2D \ac{IR} studies suggested the Zundel form to  dominate the  proton-transfer
spectroscopic signature in bulk  water \cite{Thaemer2015, Dahms2017, Fournier2018, Kundu2019, Carpenter2018, Carpenter2020}.
From ab initio molecular-dynamics (AIMD) work
it was concluded that an excess proton in bulk  liquid water is predominantly
 present in the  Eigen state  and that the Zundel form plays the role of a relatively short-lived transfer or barrier  state
  \cite{Marx1999,Napoli2018, Roy2020}.
As the separation between the  two water oxygen atoms  that flank the excess proton decreases,
the relative stability  changes and
the Zundel form becomes eventually  preferred over the Eigen form \cite{Komatsuzaki1994},
 it transpires that excess proton and water motion are dynamically coupled.
 As a consequence,
proton transfer from one water molecule to a neighboring one
not  only involves motion of the proton  but  also of the flanking water molecules and even
further water neighbors,  making the kinetics  highly collective
\cite{Kulig2013, Biswas2017, Daly2017, Wang2017, Esser2018, Napoli2018, Kundu2019, Fischer2019, Carpenter2020, Calio2021}.

In isolated  H$_5$O$_2{}^{+}$  clusters  and  protonated water wires the situation is different from bulk:
 Experimental \cite{Asmis2003a,Headrick2005,Guasco2011,Dahms2016,Dahms2017} and theoretical \cite{Tuckerman1997,Sauer2005a,Vendrell2007a,Agostini2011a,Kulig2013,
 Marsalek2016, Biswas2017,Daldrop2018a}
work  demonstrated that the oxygen-oxygen distance is decreased and  the Zundel form is more stable than the hydronium form. By chemical modifications of two proton acceptors in gas-phase clusters,  proton-transfer energy barriers of variable heights could be  demonstrated  \cite{Wolke2016}.
 Proton-transfer barriers also exist inside proteins, where amino-acid side chains that act as proton donors can be  located
at variable separations \cite{Wolf2010, Tripathi2019, Friedrich2020, Yang2022}.
Thus,  energetic barriers for proton transfer exist in a variety of systems and produce characteristic spectroscopic signatures that fundamentally go beyond the established normal-mode picture, as we show in this paper.

The excess proton has a high net charge and  during a transfer event
covers significant distances over short times, consequently, \ac{IR}  linear and non-linear spectroscopy are very suitable methods to detect proton-transfer events and have been applied to bulk acidic solutions \cite{Zundel1968},
acidic water clusters \cite{Asmis2003a},
gas-phase ions \cite{Saykally1988} and
proteins \cite{Barth2007}.
Interpretation of experimental spectra is traditionally based on normal-mode analysis around one or multiple local energy minima,
where the normal-mode frequency $f_{\rm NM}$ defines a vibrational time scale according to $\tau_{\rm NM} = 1/f_{\rm NM}$. %
But if a barrier exists, two additional time scales emerge, the
transfer-waiting time  $\tau_{\rm TW}$, which is the time the proton waits in one minimum before it transfers
\cite{Kramers1940, Williams1972, Kappler2018},  and the
transfer-path (TP) time
$\tau_{\rm TP}$, which is the time the  actual transfer  over the barrier takes
\cite{Hummer2004,Faccioli2006,Chung2009,Kim2015,Cossio2018}.

In this paper,
we  show by a combination of stochastic  theory and \ac{AIMD} simulations,
that the normal-mode, the transfer-waiting and the TP  time scales, which together
 characterize the transfer-waiting kinetics, leave distinct and characteristic spectroscopic traces.
 As a specific example,
we consider  a H$_5$O$_2{}^+$ cation  in gas phase.
In order to probe different  proton-transfer barrier heights, we  constrain the separation
between the two water oxygen atoms at variable fixed distances,
applicable to proteins and other systems where proton accepting residues
are  positioned at well-defined distances \cite{Wolf2010, Tripathi2019, Friedrich2020}.
While   the transfer-waiting time depends exponentially on the barrier height $U_0$ as
 $\tau_{\rm TW} \sim e^{U_0/k_BT}$ \cite{Kramers1940, Kappler2018},
the normal-mode time scale $\tau_{\rm NM}$ is determined
by  the stiffness of the effective harmonic potential  $k$ and the effective mass $m$ according to
 $\tau_{\rm NM} = 2 \pi \sqrt{m/k} \sim 1/\sqrt{U_0}$,
and  the TP time depends logarithmically on $U_0$ as
 $\tau_{\rm TP} \sim \ln (U_0/k_BT)/U_0$ \cite{Chung2009,Kim2015,Cossio2018}.
From the different functional dependencies  on $U_0$,
one expects  for not too low barrier heights
$\tau_{\rm NM}  \sim  \tau_{\rm TP} < \tau_{\rm TW}$.
Indeed, for an oxygen-oxygen distance of $R_{\mathrm{OO}}=\SI{2.64}{\ang}$, which  in our \ac{AIMD} simulations
of the H$_5$O$_2{}^+$ cation leads to
a moderate  effective barrier height of $U_0=2.0\,k_BT$,
the normal-mode spectroscopic contributions  lie between \SIrange{1000}{2000}{cm^{-1}},
 the  TP  contribution
 turns out to  be a  rather well defined band  centered around \SI{800}{cm^{-1}},
and since the  waiting-time distribution is rather broad, the transfer-waiting  contribution
 forms a  continuum band below \SI{500}{cm^{-1}}
that reaches deep  into the GHz range, in agreement with experimental THz absorption
measurements \cite{Decka2015, Brunig2022b}.

Our \ac{AIMD} results show that
the broad low-frequency transfer-waiting spectral contribution crucially depends on the barrier height,
controlled by the relative distance of the water molecules sharing the excess proton.
In contrast, the TP spectral contribution shifts only slightly with barrier height,
in agreement with transfer kinetic  theory  \cite{Chung2009,Kim2015,Cossio2018}.
Isotope exchange of the excess proton  on the other hand affects the TP contribution but not the waiting-time contribution, as we predict by stochastic theory.
  In summary, we show that the spectroscopic signature of proton barrier crossing reflects transfer-waiting statistics as well as TP kinetics and in particular cannot be modeled by a succession of normal modes located across the barrier.
Our results  also apply to experimental systems with fluctuating barrier heights, such as acidic water, as recently considered by a combined theoretical/experimental study \cite{Brunig2022b}:
We show that the spectrum of  unconstrained H$_5$O$_2{}^+$  can be quite accurately reproduced by Boltzmann averaging of spectra of constrained systems, thus all features we see in our constrained simulations are also expected in experimental systems where the proton acceptor separation can fluctuate.
Quantum zero-point-motion effects reduce the effective  barrier height
\cite{Tuckerman1997,Marx1999,Napoli2018,Calio2021,Schran2019},
 but for large enough barrier heights
are not expected to eliminate the spectroscopic  features we predict, as discussed in SI section \ref{qmBarrierSection}.

\section{Results and Discussion}

We perform \ac{AIMD} simulations of a single H$_5$O$_2{}^{+}$  cation
with a total trajectory length of 5 ns
for several constrained oxygen separations as well as for unconstrained oxygens  (see Methods for details).
Suitable reaction coordinates are the oxygen-oxygen distance $R_{\mathrm{OO}}$
and the excess-proton distance from the  oxygen mid-point position,
$d=\frac{1}{2}(R_{\mathrm{O}_{1}\mathrm{H}}-R_{\mathrm{O}_{2}\mathrm{H}})_x$,
projected onto the x-axis that connects the two oxygens, as illustrated in fig.~\ref{zundel_intro}A.
 %The sign of $d$ therefore indicates an excess-proton position near one or the other water molecule.
The two-dimensional  free energy in fig.~\ref{zundel_intro}D, calculated from the
probability distribution of unconstrained simulations according to
$U(\mathrm{R}_{\mathrm{OO}}, d) = -k_B T \ln p(\mathrm{R}_{\mathrm{OO}}, d)$,
 demonstrates that the global minimum of the free energy
 is located around $R_{\mathrm{OO}}=\SI{2.40}{\ang}$ and $d=0$.
 This is the symmetric Zundel state, where the excess proton is symmetrically shared by the oxygens \cite{Zundel1968}.
 For $R_{\mathrm{OO}} >\SI{2.55}{\ang}$ a  double-well free-energy landscape along $d$ appears,
which  indicates  a preferred localization of the excess proton near one  water molecule,
analogous to the Eigen state  in bulk water \cite{Wicke1954}.
The  excess proton trajectory for constrained $R_{\mathrm{OO}}=\SI{2.64}{\ang}$  in fig.~\ref{zundel_intro}B
is typical for the thermally activated barrier crossing of a weakly damped massive particle     \cite{Kappler2018}
and involves   a moderate barrier height of
$U_0=2.0\,k_B T$,  as seen in the   corresponding  free-energy profile  in fig.~\ref{zundel_intro}C.
Most of the time the excess proton is   part of a H$_3$O${}^+$  molecule
and vibrates in one of the two free-energy minima with an oscillation time described by the normal-mode time $\tau_{\rm NM} = \SI{17}{fs}$ (inset fig.~\ref{zundel_intro}B),
while from time to time the proton  suddenly crosses the barrier,
the mean time of such a TP  is $\tau_{\rm TP} = \SI{25}{fs}$
(inset  fig.~\ref{zundel_intro}B). The longest time scale
 is the transfer-waiting  time, which for $R_{\mathrm{OO}}=\SI{2.64}{\ang}$ is  $\tau_{\rm TW} = 440$ fs.
In fig.~\ref{zundel_intro}E we show as a grey solid line the absorption spectrum of the unconstrained
H$_5$O$_2{}^{+}$  cation along $x$, the oxygen separation direction,
calculated from the entire nuclear and electronic polarizations (see Methods).
It  shows  in addition to the OH stretch and HOH bend bands at
3400 cm$^{-1}$ and 1800 cm$^{-1}$, respectively, a prominent feature at 1000 cm$^{-1}$, which is the Zundel normal mode,
where the excess proton vibrates in a rather soft potential produced by the two flanking water molecules
(see SI section \ref{pubDataSection} and \ref{normalModeSection} for a literature overview).
The spectrum for the constrained system with $R_{\mathrm{OO}}=\SI{2.64}{\ang}$, grey broken line, displays a band at
800 cm$^{-1}$ and a very broad shoulder that extends down to the lowest frequencies. As we  show in this paper,
these two spectral  features stem from  proton TPs and proton transfer-waiting-time  stochastics, respectively,
 and are the only
spectroscopic contributions that reflect  the actual proton-transfer kinetics.
Interestingly, the spectral contribution of only the excess proton
for fixed $R_{\mathrm{OO}}=\SI{2.64}{\ang}$ (blue solid line, multiplied by a factor of 2)
is almost identical to the full spectrum  (grey line), so we conclude that the IR spectrum is predominantly caused by proton motion and can thus be used to investigate  excess-proton dynamics
(more details are given in SI section \ref{hDecompSection}).
In fact, the spectrum of the unconstrained system (black broken line)
agrees well with the  free-energy-weighted
Boltzmann average over  constrained spectra with  different $R_{\mathrm{OO}}$ values
(black dashed line, see SI section \ref{dichroicSpectraSection} for details),
indicating that the proton and the oxygen dynamics decouple.
Our simulation model with constrained oxygen-oxygen separation thus  is
also a tool to decompose and thereby understand unconstrained system dynamics
(a finding that is obvious only for static observables  \cite{Sprik1998}).

\begin{figure*}
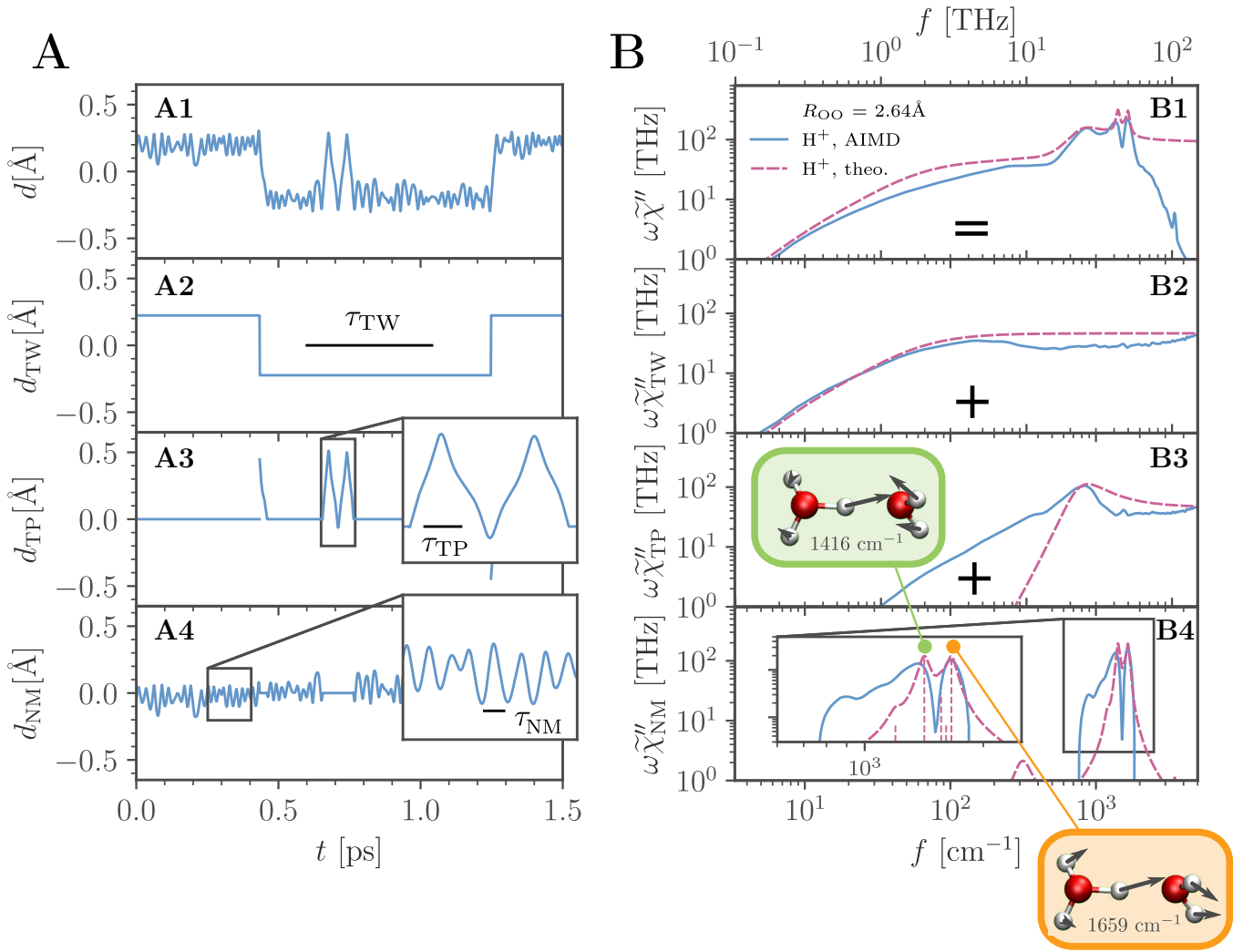

\centering
  \begin{overpic}[width=\textwidth]{{/../figs/zundel_decomp_2}.eps}
\put(2,72){\huge \bf A}
\put(12,68){\large \bf A1}
\put(12,53.5){\large \bf A2}
\put(12,39.5){\large \bf A3}
\put(12,25.5){\large \bf A4}
\put(49,72){\huge \bf B}
\put(93.,68){\large \bf B1}
\put(93.,53.5){\large \bf B2}
\put(93.,39.5){\large \bf B3}
\put(93.5,25.5){\large \bf B4}
\end{overpic}
\caption{\ac{AIMD} simulations of a
H$_5$O$_2{}^{+}$  cation
 with constrained $R_{\mathrm{OO}}=\SI{2.64}{\ang}$. \textbf{A}   Decomposition of the
  excess-proton trajectory $d(t)$ into the two-state  transfer-waiting contribution $d_{\mathrm{TW}}(t)$,
  the TP contribution $d_{\mathrm{TP}}(t)$ and the remaining normal-mode contribution $d_{\mathrm{NM}}(t)$.
  \textbf{B} Blue solid lines show the simulated  excess-proton spectrum  $\omega \widetilde \chi''$ and its decomposition into the
  transfer-waiting  $\omega \widetilde \chi''_{\mathrm{TW}}$, the  TP $\omega \widetilde \chi''_{\mathrm{TP}}$
  and the normal-mode contribution $\omega \widetilde \chi''_{\mathrm{NM}}$. The red broken lines in B2 and B3
  show the corresponding theoretical predictions
 according to  eqs.~\eqref{BarrierCrossingSpectrum} and \eqref{TransferPathSpectrum}.
The red broken line in B4 shows the  normal-mode spectrum
including   friction-induced line broadening.
The snapshots illustrate the two dominant normal modes  at \SI{1416}{cm^{-1}} and \SI{1659}{cm^{-1}}.}
\label{decomp}
\end{figure*}

In order to distinguish  transfer-waiting, TP and normal-mode spectral contributions,
the proton trajectory $d(t)$ is decomposed
 according to $d(t)=d_{\mathrm{TW}}(t)+d_{\mathrm{TP}}(t)+d_{\mathrm{NM}}(t)$,
 as illustrated in fig.~\ref{decomp}A for $R_{\mathrm{OO}}=\SI{2.64}{\ang}$.
 The transfer-waiting part $d_{\mathrm{TW}}(t)$ describes  two-state kinetics  with instantaneous transfers
  when the trajectory last crosses a free-energy minimum  at $d^*_{\mathrm{TW}}= \pm\SI{0.22}{\ang}$.
 The TP contribution $d_{\mathrm{TP}}(t)$ consists of   transfer trajectories between
 last and first crossing the  free-energy minima, including recrossings
 where the proton shuttles repeatedly back and forth between the minima.
 Recrossings are rather frequent for  the low friction experienced by the proton
   \cite{Kappler2018} (see SI section \ref{recrossingSection}),  a three-fold recrossing event  is
 seen in the proton trajectory in fig.~\ref{decomp}A at $t=\SI{0.6}{ps}$.
 Finally,  the normal-mode part $d_{\mathrm{NM}}(t)$ comprises  the trajectory remainder.

 \begin{figure*}[tb]
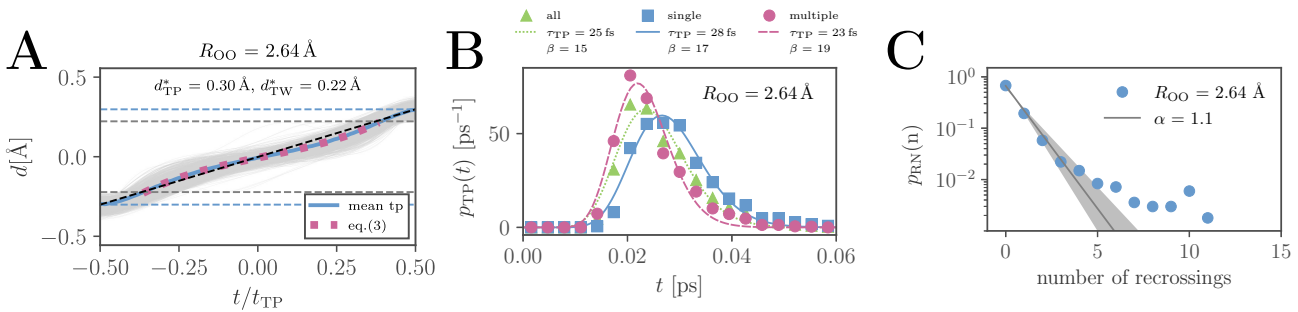

\centering
\begin{overpic}[width=0.32\textwidth]{{/../figs3/tp/zundel_tpFitMinMaxAvgD_d2.64_wBG}.eps}
\put(0,56){\huge \bf A}
\end{overpic}
\begin{overpic}[width=0.32\textwidth]{{/../figs3/tp/zundel_tp_d2.64_MinMaxTimes_ExpN}.eps}
\put(0,56){\huge \bf B}
\end{overpic}
\begin{overpic}[width=0.32\textwidth]{{/../figs3/tp/zundel_tp_recrStat_expFit_d2.64}.eps}
\put(0,56){\huge \bf C}
\end{overpic}
\caption{TP statistics. {\bf A} Ensemble of all 2829  proton  TPs for constrained $R_{\mathrm{OO}}=\SI{2.64}{\ang}$
(grey lines) as a function of the rescaled time  $t/t_{\mathrm{TP}}$, where $t_{\mathrm{TP}}$ is the individual TP  time.
Horizontal blue dashed lines indicate the mean
TP  terminal positions $\pm d^*_{\mathrm{TP}}/2$, defined by the TP  turning points,
while horizontal grey dashed lines indicate the free-energy minima $\pm d^*_{\mathrm{TW}}/2$.
The simulated mean TP  (blue line) agrees well  with the path-integral prediction  eq.~\eqref{eq:tpPathTheo} (red dotted line).
The straight black dashed line approximates the TP shape also quite well.
{\bf B}  Distribution $p_{\mathrm{TP}}$  of all TPs  (green triangles) and  a decomposition into
 single  (non-recrossing, blue squares) and multiple (recrossing, red dots) TPs  together
 with fits according to eq. (\ref{pTP}).
 {\bf C} Recrossing-number probability distribution $p_{\mathrm{RN}}(n)$ compared  to an exponential fit $p_{\mathrm{RN}}(n)=(1-  e^{- \alpha })  e^{- \alpha n}$,  the confidence interval $\alpha \pm 20\%$ is shown by grey lines.}
\label{tps}
\end{figure*}

Fig.~\ref{decomp}B shows in blue the simulated excess-proton spectrum decomposed into its three components according to
$\widetilde \chi'' = \widetilde \chi''_{\mathrm{B}} + \widetilde \chi''_{\mathrm{TP}} + \widetilde \chi''_{\mathrm{NM}}$,
the red broken lines show  theoretical predictions (which will be explained further below).
Trajectory decomposition in  the time domain creates spectral cross contributions, which  are relatively small,
as shown in SI sections \ref{linearResponseSection} and \ref{crossContribSection},
 and are added to $\widetilde \chi''_{\mathrm{NM}}$.
The transfer-waiting spectrum $\widetilde \chi''_{\mathrm{TW}}$ in fig.~\ref{decomp}B2 displays a pronounced
low-frequency shoulder, which reflects the  transfer-waiting-time distribution.
The TP spectrum  $\widetilde \chi''_{\mathrm{TP}}$ in fig.~\ref{decomp}B3
is a rather well defined band  at \SI{800}{cm^{-1}}.
Even though the time  fraction the excess proton spends on TPs  is only 16\% for $R_{\mathrm{OO}}=\SI{2.64}{\ang}$,
the spectral contribution is significant due to the large and quick charge displacement: The proton transfer
 velocity of roughly
$v_{\rm TP} = 2d^*_{\rm TP}/\tau_{\rm TP} = \SI{0.44}{\ang}/ \SI{25}{fs} = \SI{1.8e3}{m/s}$ is
  slightly larger than the proton thermal velocity
of $v_{\rm th} = \sqrt{k_BT/m_p} = \SI{1.5e3}{m/s}$, where $m_p= 1.7 \times 10^{-27}$ kg is the proton mass.
This confirms previous findings that TPs correspond to the high-energetic part of the Maxwell-Boltzmann ensemble,
i.e. the excess proton initiates  a TP only when its kinetic energy is above  average \cite{Daldrop2016}.
The normal-mode spectrum  $\widetilde \chi''_{\mathrm{NM}}$ in fig.~\ref{decomp}B4 consists of two main peaks.

\begin{figure*}[tb]
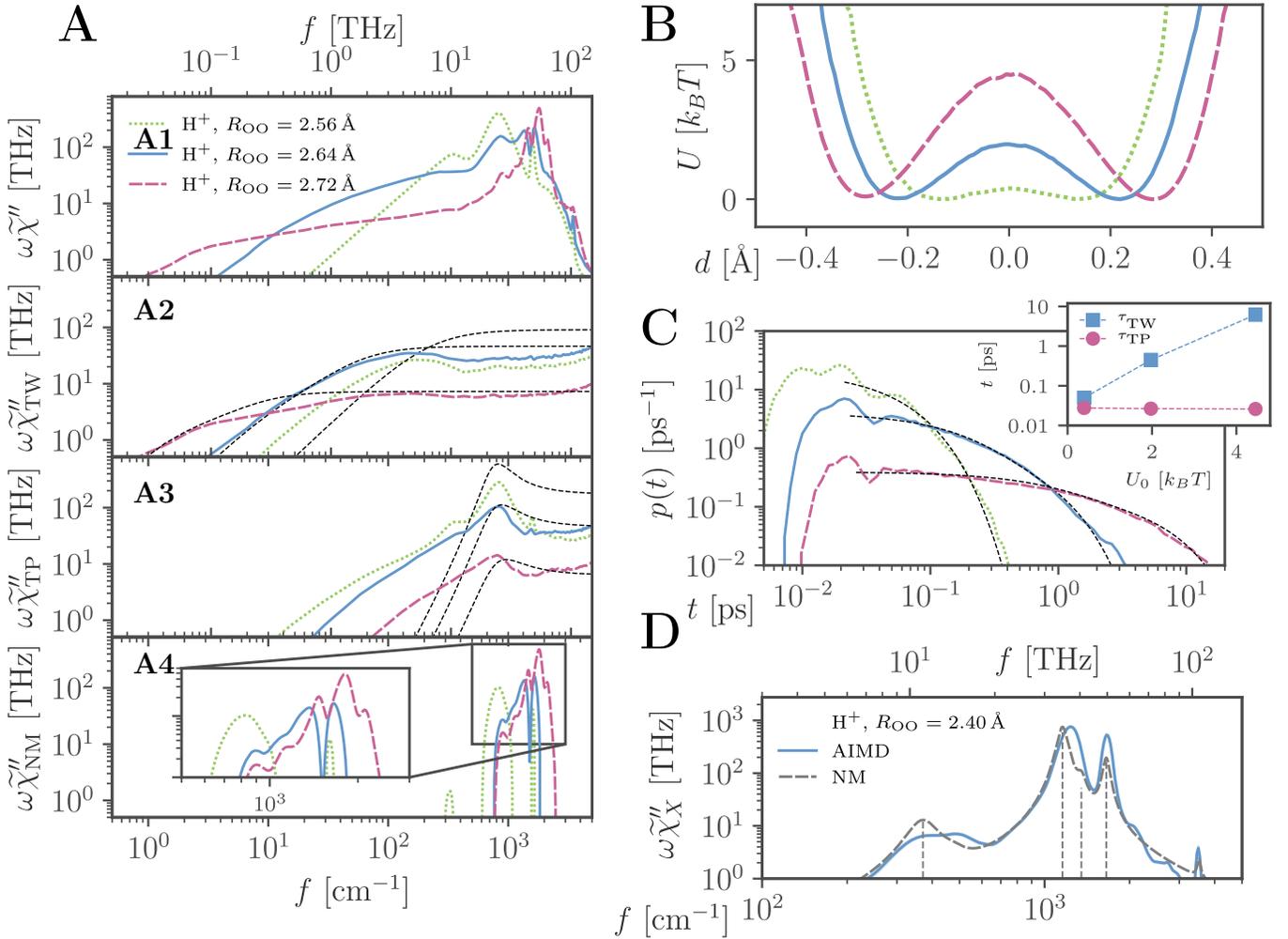

\centering
\begin{overpic}[width=\textwidth]{{/../figs/zundel_systems}.eps}
\put(4,71){\huge \bf A}
\put(10,62.5){\large \bf A1}
\put(10,49){\large \bf A2}
\put(10,34.5){\large \bf A3}
\put(10,21){\large \bf A4}

\put(50,71){\huge \bf B}
\put(50,47){\huge \bf C}
\put(50,23){\huge \bf D}
\end{overpic}
\caption{Decomposition of the excess-proton spectra for various constrained $R_{\mathrm{OO}}$. \textbf{A} \ac{AIMD} spectra are shown as colored lines and theoretical predictions
  are shown for the transfer-waiting contribution, eq.~\eqref{BarrierCrossingSpectrum},
  in A2
  and for the TP contribution, eq.~\eqref{TransferPathSpectrum}, in A3 as thin black broken lines.
 See  SI section \ref{systemsSection} for details.
    \textbf{B} Proton free energies landscapes extracted from simulation trajectories.
  \textbf{C} Transfer-waiting-time distributions together with bi-exponential fits
  (black broken lines).
  The inset shows the mean  transfer-waiting times  $\tau_{\mathrm{TW}}$  and the mean TP times
   $\tau_{\mathrm{TP}}$  as a function of the free-energy barrier height $U_0$.
   \textbf{D} \ac{IR} spectrum of the excess proton (blue solid line) in the H$_5$O$_2{}^{+}$ cation with fixed $R_{\mathrm{OO}}=\SI{2.40}{\ang}$ compared to the normal-mode spectrum  including frictional   line-broadening (grey broken line).
   Vertical grey broken lines denote the dominant normal modes.}
\label{systems}
\end{figure*}

 We will now present  analytic theories for each simulated spectral contribution shown in  figs.~\ref{decomp}B2--B4.
 A stochastic  two-state process  has the spectrum
\begin{align}
\omega \widetilde \chi_{\mathrm{TW}}''(\omega) = \frac{2q^2{d^*_{\mathrm{TW}}}^2}{V \epsilon_0 k_BT}\ \text{Re}\left(\frac{\omega^2 \tilde q_{\mathrm{TW}}(\omega)}{1 - \tilde p_{\mathrm{TW}}(\omega)^2}\right)
\label{BarrierCrossingSpectrum}
\end{align}
and depends on the Fourier-transformed transfer-waiting-time distribution $\tilde p_{\mathrm{TW}}(\omega)$ and
the survival distribution $\tilde q_{\rm TW}(\omega)$,
which is defined as
$q_{\rm TW}(t) = \int_t^{\infty} p_{\mathrm{TW}}(t') dt'$,  the  positions of the free-energy minima   $\pm d^*_{\mathrm{TW}}$,
the excess proton charge $q=e$ and the system volume $V$  (see SI Sect.  \ref{binaryResponseSection} for a detailed derivation).
Using $d^*_{\mathrm{TW}}=\SI{0.22}{\ang}$  and bi-exponential fits
for $p_{\mathrm{TW}}(t)$  to the simulation data in fig.~\ref{systems}C,
$\omega \widetilde \chi''_{\mathrm{TW}}(\omega)$ according to eq.~\eqref{BarrierCrossingSpectrum} (red broken line)
matches  the simulation data (blue solid line)  in fig.~\ref{decomp}B2 very well without any  fitting parameters.
For a single-exponential waiting-time  distribution,
$p_{\mathrm{TW}} (t)=\tau_{\mathrm{TW}}^{-1}\exp(-t/\tau_{\mathrm{TW}})$,
eq.~\eqref{BarrierCrossingSpectrum} simplifies  to
\begin{align}
\omega \widetilde \chi_{\mathrm{TW}}''(\omega) = \frac{2 q^2 {d^*_{\mathrm{TW}}}^2 }{V\epsilon_0 k_BT} \frac{\tau_{\mathrm{TW}} \omega^2 }{(4 + \tau_{\mathrm{TW}}^2 \omega^2)},
\label{ExpBarrierCrossingSpectrum}
\end{align}
which shows that the  spectrum  is identical to an overdamped harmonic
  oscillator with  a corner frequency  $\omega_{\mathrm{TW}}^* \sim 1/ \tau_{\mathrm{TW}}$
  (see SI section \ref{hoResponseSection} for details).
  For large frequencies $\omega \widetilde \chi_{\mathrm{TW}}''$
  is constant and proportional to  the transfer-waiting rate,
  $\omega \widetilde \chi_{\mathrm{TW}}'' \sim 1/ \tau_{\mathrm{TW}}$,
  for small frequencies  $\omega \widetilde \chi_{\mathrm{TW}}'' \sim  \tau_{\mathrm{TW}}\,\omega^2$.

 The TP spectral contribution depends on  the TP shape.
The ensemble of all 2829  TPs observed in the simulations
 for $R_{\mathrm{OO}}=\SI{2.64}{\ang}$ is shown in fig.~\ref{tps}A
(grey lines),   together  with the mean TP  (blue solid line) obtained by position averaging.
The path-integral saddle-point prediction for the TP  shape over a parabolic barrier  \cite{Cossio2018},
\begin{align}
 d_\text{TP}(t)={d^*_{\mathrm{TW}}}
 \left[ e^{ t / \kappa} - e^{-t / \kappa} \right]
 /{\cal N},
 \label{eq:tpPathTheo}
\end{align}
 (red dotted line) matches the simulated mean TP shape very well
(${\cal N}$ is a normalization constant).
In  SI section \ref{tpPathIntegralSection} it is shown that eq.~\eqref{eq:tpPathTheo} corresponds
 to  the exact mean TP shape  in the high-barrier limit \cite{Kim2015}.
 The fitted  characteristic time    $\kappa = {d^*_{\mathrm{TW}}}^2\gamma/(2U_0) =\SI{6.5}{fs}$ depends on  the effective  friction coefficient $\gamma$ acting on the proton as it moves over
  the barrier.
 A straight line (black broken line)  describes the simulated mean TP shape  also quite  well.
Fig.~\ref{tps}B shows  the  TP-time  distribution of
all TPs (green triangles) together with a decomposition into
 single  (non-recrossing, blue squares) and multiple (recrossing, red dots) TPs,
 where the TP time  $\tau_{\rm TP}$  is defined from the turning points of the TPs.
 It is seen that multiple TPs that consist of recrossing trajectories are significantly faster  than
 single TPs, which reflects that recrossing protons
 have a higher  kinetic energy and thereby tend to rebounce back over the barrier.
 Fits according to the Erlang distribution \cite{Cox1977}
 \begin{equation}
 p_{\mathrm{TP}}(t)=\frac{t^{\beta-1}}{(\beta-1)!} \left(\frac{\beta}{\tau_{\rm TP}}\right)^{\beta} e^{-\beta t/\tau_{\rm TP}}
 \label{pTP}
 \end{equation}
  are shown as lines.
 In Fig.~\ref{tps}C  the simulated
recrossing-number  distribution $p_{\mathrm{RN}}(n)$ is compared
 to an exponential fit with  a decay constant  $ \alpha = 1.1$,
 40 \% of all TPs are single transfer
 events, $n=0$, while  the remaining 60 \% TPs are part of multiple events with $n>0$.

Combining the TP  time distribution $p_{\mathrm{TP}}(t)$ in the infinitely sharp limit $\beta \rightarrow \infty$,
the exponential recrossing-number  distribution $p_{\mathrm{RN}}(n)$ and approximating the TP shape as a straight line,
the  analytical result for the
TP spectral contribution (red broken line in fig.~\ref{decomp}B3) is in SI section \ref{theoryTPSection}  derived as
\begin{align}
\omega \widetilde \chi_{\mathrm{TP}}''(\omega)= &\frac{{d^*_{\rm TP}}^2q^2}{ V\epsilon_0 k_BT \tau_{\mathrm{TW}}}
\frac{64 \omega ^2 \tau _{\text{TP}}^2}{\pi^4 \left(\omega  \tau _{\text{TP}}+\pi \right){}^2} \nonumber \\
&\frac{e^{\alpha } \omega ^2 \tau _{\text{TP}}^2}{2 \cosh (\alpha ) - 2 + \left(\omega \tau _{\text{TP}} -\pi \right){}^2 }
\label{TransferPathSpectrum}
\end{align}
and matches  the simulation data (blue solid line) around the maximum quite well.
In the comparison the mean time of recrossing TPs  $\tau _{\text{TP}}=  23$ fs
from  fig.~\ref{tps}B is used, which is shown to be the dominating
time scale in  SI section \ref{theoryTPSection}. Interestingly, the TP spectrum eq.~\eqref{TransferPathSpectrum}
is a product of a Debye and a Lorentzian line shape, both with the same
characteristic frequency $f_{\text{TP}} = 1/(2 \tau _{\text{TP}})$, which explains its relative sharpness.

The  remaining normal-mode  contribution $\widetilde \chi''_{\mathrm{NM}}$ in fig.~\ref{decomp}B4
is obtained  by harmonic  analysis  of  the minimal energy structure including
line broadening from frictional damping (red broken line).
The two dominant normal modes
around  \SI{1416}{cm^{-1}} and  \SI{1659}{cm^{-1}},
which correspond to in-phase and out-of-phase coupled vibrations of the excess proton with the
hydrogens of the distant water,
are illustrated in fig.~\ref{decomp}B4
 (see Methods and SI section \ref{normalModeSection} for details).

In fig.~\ref{decomp}B1 the simulated excess-proton spectrum (blue solid line)   is compared to the sum of the theoretical
transfer-waiting, TP and normal mode predictions (red broken line),
the agreement is  good (except for very high frequencies), which
 demonstrates that eqs.~\eqref{BarrierCrossingSpectrum}
and \eqref{TransferPathSpectrum} together with the normal-mode analysis
allow to quantitatively describe  excess-proton transfer spectra.

The excess-proton spectra in fig.~\ref{systems}A1 vary significantly  for different values of  $R_{\mathrm{OO}}$.
The  excess-proton free energies from simulations  in  fig.~\ref{systems}B demonstrate that the three systems
exhibit high, moderate and low barriers.
Very pronounced  is the change of the low-frequency shoulder of the transfer-waiting contribution in fig.~\ref{systems}A2,
which moves to lower frequencies and becomes weaker with growing barrier height and is
well captured by the theoretical predictions eq.~\eqref{BarrierCrossingSpectrum}  (black broken lines)
using  bi-exponential fits to the transfer-waiting distributions in fig.~\ref{systems}C.
Eq.   \eqref{ExpBarrierCrossingSpectrum} demonstrates that the spectral differences are due to
 less frequent transfers as the barrier height increases.
The simulated mean transfer-waiting time $\tau_{\mathrm{TW}}$  in the inset of fig.~\ref{systems}C
exponentially  increases with the barrier height $U_0$,
as expected for thermally activated  barrier crossing \cite{Kramers1940, Kappler2018}.
On the other hand,  the frequency of the TP spectral contribution in fig.~\ref{systems}A3  shifts very little
 for different  $R_{\mathrm{OO}}$,
 which is well-captured by  eq.   \eqref{TransferPathSpectrum} (black broken lines)
and  reflects   the weak dependence of the TP time $\tau _{\text{TP}}$ on
 the barrier height  in the inset of fig.~\ref{systems}C,
 %For $R_{\mathrm{OO}}=\SI{2.64}{\ang}$ the transfer-path time
in agreement with the predicted
 logarithmic dependence of  $\tau _{\text{TP}}$ on the barrier height  \cite{Chung2009}.

Fig.~\ref{systems}D compares the \ac{IR} spectrum of the excess proton (blue solid line) in the H$_5$O$_2{}^{+}$
cation  to the normal-mode spectrum including frictional line broadening
(grey broken line, see SI section \ref{normalModeSection} for details)  for  fixed $R_{\mathrm{OO}}=\SI{2.40}{\ang}$,
the barrier-less global minimum of the unconstrained H$_5$O$_2{}^{+}$ cation.
The good agreement highlights that the
 barrierless Zundel state  is well described by a normal-mode analysis.
 This is in contrast to  the results for larger values of $R_{\mathrm{OO}}$ in fig.~\ref{systems}A, where
 a finite barrier exists and  the transfer-waiting and TP  spectral signatures dominate over the normal-mode contribution.

\section{Conclusions and Discussion}
In contrast to
traditional normal-mode-based  approaches to proton-transfer spectroscopy,
which consider  proton vibrations around  energy minima,
we here
investigate  the  spectrum of a proton as it actually makes the move from one energy minimum  to  another.
While the normal-mode  frequencies are on the harmonic-approximation level determined by the curvature of the energy landscape and by the effective mass, two fundamentally different time scales govern the
barrier-crossing absorption spectrum:
the mean time the proton waits in a potential minimum before it crosses the barrier, the transfer-waiting
 time, and the mean time it takes the proton to actually move over the barrier once it has left the potential minimum,
 the so-called transfer-path (TP)  time.
While the TP
  time distribution is rather narrow, which leads to a well-defined TP band, the transfer-waiting
 times are  broadly distributed, which leads to a wide  spectral absorption down to low frequencies.
 Recent experimental studies on hydrochloric acid solutions  in the THz regime indeed observed broad absorption that by comparison with AIMD simulations could be attributed to proton motion \cite{Decka2015,Brunig2022b}.

The \ac{AIMD} simulations  of single H$_5$O$_2{}^+$ cations reveal a high similarity of
  excess-proton-only spectra and  spectra from all nuclei and electronic polarizations.
  This emphasizes the impact of proton-transfer processes on experimentally measured spectra
  and allows in turn to develop a stochastic spectral theory based on excess-proton motion only.
  The excess-proton transfer between two water molecules depends strongly on the separation of the two water oxygens. For oxygen-oxygen separations $R_{\mathrm{OO}} \geq \SI{2.5}{\ang}$ a barrier crossing is involved, whereas for closer separations the proton is rather located directly in between the two water molecules.

An H/D isotope exchange of the excess proton does not shift the low-frequency transfer-waiting signature,
 as shown in SI section \ref{deuteronSpectraSection}, which is expected since
 the  excess-proton barrier crossing is a  friction-dominated process
 and mass plays only a minor role,  as discussed in SI section \ref{deuteronRatesSection}.
In contrast, TP  and normal-mode signatures show isotope effects,
which suggests how to experimentally  distinguish barrier crossing from the other  spectral contributions.
For the normal-mode spectral contribution the isotope effect is well known (see section \ref{hoResponseSection} in the SI),
 the mass-dependence of the TP spectral contribution is rather subtle and depends on the stochastic  mass-friction
 balance  (see section \ref{deuteronRatesSection} in the SI).

The spectroscopic signatures of proton transfer  are most pronounced along  the transfer direction,
as shown in section \ref{dichroicSpectraSection} in the SI, thus dichroic measurements  \cite{Daldrop2018a, Yang2022} are most suitable to observe these features.

\subsection*{Methods}

The Born-Oppenheimer \ac{AIMD} simulations of the H$_5$O$_2{}^+$ cation were performed with the CP2K 4.1 software package  using a doubly polarizable triple-zeta basis set for the valence electrons, dual-space pseudopotentials, the BLYP exchange-correlation functional and D3 dispersion correction~\cite{Hutter2014, Kendall1992,Grimme2010}. The simulation box size was
 $10\times 10\times \SI{10}{\ang^3}$ and the cutoff for the plane-wave representation  \SI{400}{Ry}.
For each constrained system \SI{20}{ps} simulations with a time step of \SI{0.5}{fs} were performed under NVT conditions at \SI{300}{K} by coupling all atoms to a CSVR thermostat with a time constant of \SI{100}{fs}, which has been shown to be exceptionally good for preserving vibrational dynamics \cite{Bussi2007}. Consequently a number of independent simulations with a time step of \SI{0.25}{fs} were performed under NVE conditions starting from different snapshots of the NVT data, $12 \times \SI{20}{ps}$ for the systems with $R_{\mathrm{OO}}\leq\SI{2.5}{\ang}$ and $\geq 20 \times \SI{60}{ps}$ for the systems with $R_{\mathrm{OO}}\geq\SI{2.5}{\ang}, 20 \times \SI{90}{ps}$ for $R_{\mathrm{OO}}=\SI{2.72}{\ang}$. Even though the time step was chosen very small, some systems did not preserve energy during the NVE simulation due to unfavorable starting conditions and the small number of degrees of freedom. These systems were excluded from further analysis. The data of  systems with constrained oxygen atoms stem from  NVE simulations, totaling in \SIrange{240}{1800}{ps} simulation time for each system.
In case of the unconstrained system, the oxygen atoms were only constrained in the $yz$-plane.
Nevertheless the NVE simulations were less stable due to large spatial fluctuations along $x$. For this system
 NVT simulations with a total simulation time of \SI{20}{ps} were performed.

Linear response theory relates the dielectric susceptibility $\chi(t)$ to the equilibrium autocorrelation of the dipole moment $C(t)=\langle \bm{p} (t)\bm{p}(0)\rangle$, reading in Fourier space
\begin{align}
\label{linearResponseFT}
\widetilde \chi(\omega) = \frac{1}{V \epsilon_0 k_BT}\left( C(0) - i \frac{\omega}{2} \widetilde C^+(\omega) \right),
\end{align}
with system volume $V$, thermal energy $k_BT$ and vacuum permittivity $\epsilon_0$. \ac{IR} spectra can therefore be calculated straight-forwardly from sufficiently sampled trajectories of the \ac{AIMD} simulation data using eq.~\eqref{linearResponseFT} and the Wiener-Kintchin relation, derived in SI section \ref{WienerKintchineSection}.
Quantum corrections have previously been addressed  \cite{Ramirez2004},
but were not applied here. The dipole moments were obtained after Wannier-center localization of the electron density at a time resolution of \SI{2}{fs}. The power spectra were smoothed using Gaussian kernels with widths that are logarithmically increasing from \SI{20}{cm^{-1}} centered at \SI{20}{cm^{-1}} to \SI{100}{cm^{-1}} centered at \SI{5000}{cm^{-1}}.
All presented spectra were scaled by the volume of two water molecules, $V=\SI{0.060}{nm^3}$,
which follows from the density of water at atmospheric pressure and \SI{300}{K},  $\rho = \SI{0.99}{g/ml}$.
The normal-mode analysis was performed using the implementation in CP2K 4.1 by diagonalizing the Hessian of energetically optimal structures for the same system parameters as in the \ac{AIMD} simulations.
The normal modes were obtained as the Eigenvectors of the Hessian, the Eigenvalues are the frequencies.
A projection of the Eigenvectors onto the excess-proton coordinate gave their spectral contributions.
Line broadening resulted from frictional damping
with the same fitted  friction coefficient $\gamma=\SI{16}{u/ps}$ for all normal modes
(see SI Sect. \ref{hoResponseSection} for details).

\section*{Supplementary Material}

See Supplementary Material for detailed derivations, analysis procedures, additional data and discussion

\section*{Author contributions}
F.N.B. and R.R.N. conceived the theory and designed the simulations. F.N.B. performed the \ac{AIMD} simulations and analyzed the data. P.H. performed the quantum-mechanical zero-point calculations. W.K.K. contributed to the transfer-path-shape theory. All authors discussed the results, analyses and interpretations. F.N.B. and R.R.N. wrote the paper with input from all authors.

\begin{acknowledgments}
We gratefully acknowledge support by the DFG grant SFB 1078, project C1 and computing time on the HPC clusters at the physics department and ZEDAT, FU Berlin. W.K.K. acknowledges the support by a KIAS Individual Grant (CG076001) at Korea Institute for Advanced Study.
\end{acknowledgments}

\section*{Competing interests}
The authors declare no competing interests.

\section*{Data Availability Statement}
The data that support the findings of this study are available from the corresponding author upon request.

\begin{acronym}[Bash]
  \acro{AIMD}{ab initio molecular-dynamics}
  \acro{DFT}{density functional theory}
  \acro{FPT}{first-passage time}
  \acro{GLE}{generalized Langevin equation}
  \acro{GH}{Grote-Hynes}
  \acro{IR}{infrared}
  \acro{LE}{Langevin equation}
  \acro{MD}{molecular dynamics}
  \acro{MFPT}{mean first-passage time}
  \acro{MFP}{mean first-passage}
  \acro{MSD}{mean squared displacement}
  \acro{NM}{normal-mode}
  \acro{PTP}{$p(\text{TP}|q)$}
  \acro{PME}{particle mesh Ewald~\cite{pronk2013gromacs}}
  \acro{PMF}{potential of mean force}
  \acro{PGH}{Pollak-Grabert-Hanggi}
  \acro{RC}{reaction coordinate}
  \acro{RDF}{radial distribution function}
  \acro{RTT}{round-trip time}
  \acro{TP}{transfer path}
 \end{acronym}
 
 % Create the reference section using BibTeX:
\bibliography{bibliography.bib}

\end{document}

\title{Supplementary information:\\ Proton-transfer spectroscopy beyond the normal-mode scenario}

\author{Florian N. Br\"unig}
\affiliation{Freie Universität Berlin, Department of Physics, 14195 Berlin, Germany}

\author{Paul Hillmann}
\affiliation{Freie Universität Berlin, Department of Physics, 14195 Berlin, Germany}

\author{Won Kyu Kim}
\affiliation{Korea Institute for Advanced Study, School of Computational Sciences, Seoul 02455, Republic of Korea}

\author{Jan O. Daldrop}
\affiliation{Freie Universität Berlin, Department of Physics, 14195 Berlin, Germany}

\author{Roland R. Netz}
\email[]{rnetz@physik.fu-berlin.de}
\affiliation{Freie Universität Berlin, Department of Physics, 14195 Berlin, Germany}

\date{\today}

\pacs{}

\maketitle

\section{Quantum zero-point motion effects}
\label{qmBarrierSection}

Quantum zero-point motion  is known to smear out particle distributions and thereby to increase the particle density at barriers. Therefore in this section, we estimate the potential barrier height of a double-well potential for which the ground-state probability density develops a single minimum.

\begin{figure}[h]
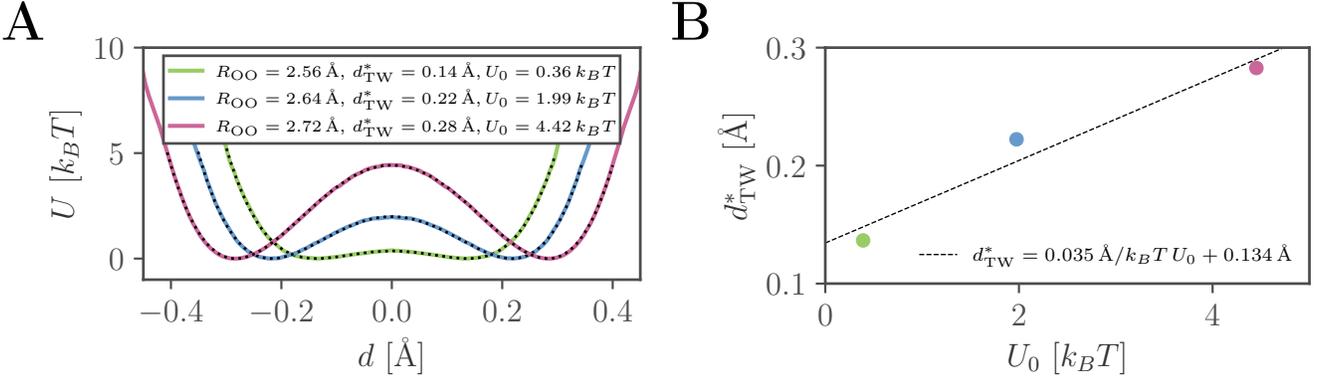

  \centering
  \begin{overpic}[width=.49\textwidth]{{/../figs3/fe/zundel_ffit_dw}.pdf}
  \put(-2,56){\huge \bf A}
\end{overpic}
  \begin{overpic}[width=.49\textwidth]{{/../figs3/fe/zundel_doverU}.pdf}
  \put(-2,56){\huge \bf B}
\end{overpic}
  \caption{{\bf A} Free-energy profiles $U(d)$ for fixed $R_{\mathrm{OO}}$, extracted from constrained ab initio molecular-dynamics (AIMD) simulations of the H$_5$O$_2{}^{+}$ cation, are shown as colored lines.
  Fits according to eq.~\eqref{eq:dw} are shown as black dotted lines with fit parameters given in the legend. {\bf B} Dependence of $d^*_{\rm TW}$ on $U_0$. Here a linear fit is shown as a black dashed line.
\label{dwFits}}
\end{figure}

The one-dimensional stationary Schr\"odinger equation
\begin{align}
-\frac{\hbar^2}{2m}\frac{\partial^2}{{\partial d}^2} \phi(d) &= (U(d) - E) \phi(d)
\end{align}
is solved numerically for the quartic double-well potential
\begin{align}
\label{eq:dw}
  U(d) &= U_0 \left(\left(\frac{d}{d_{\rm TW}^*}\right)^2 -1\right)^2,
\end{align}

which is shown in fig.~\ref{dwFits}A to approximate the effective potential describing  the excess proton distribution obtained from our \ac{AIMD} simulations very well.

The  equations are rescaled as

\begin{align}
  \label{eqn:SE}
  \frac{\partial^2}{{\partial \tilde  x}^2} \phi(\tilde x)
  &= (\widetilde U(\tilde x) - \widetilde E) \phi(\tilde x),\\
 \widetilde V(\tilde x) &= \widetilde U_0 (\tilde x^2 -1)^2,
\end{align}

with $\tilde x=d/d_{\rm TW}^*$, $\widetilde U_0=2mU_0 {d_{\rm TW}^*}^2/\hbar^2$ and
$\widetilde E=2m E {d_{\rm TW}^*}^2/\hbar^2$. Using the proton mass $m=\SI{1.7e-27}{kg}$, $\hbar=\SI{1.1e-34}{Js}$ and $d^*_{\rm TW}=\SI{0.22}{\ang}$ for the system with $R_{\mathrm{OO}}=\SI{2.64}{\ang}$, the rescaling factor for the energies is $\hbar^2/(2m {d_{\rm TW}^*}^2)=\SI{7.4e-21}{J}=1.8\ k_BT$ at \SI{300}{K}. Note that this scaling factor implicitly depends on $U_0$ due to the dependence of $d^*_{\rm TW}$ on $U_0$ shown in fig.~\ref{dwFits}B.

\begin{figure}[tbh]
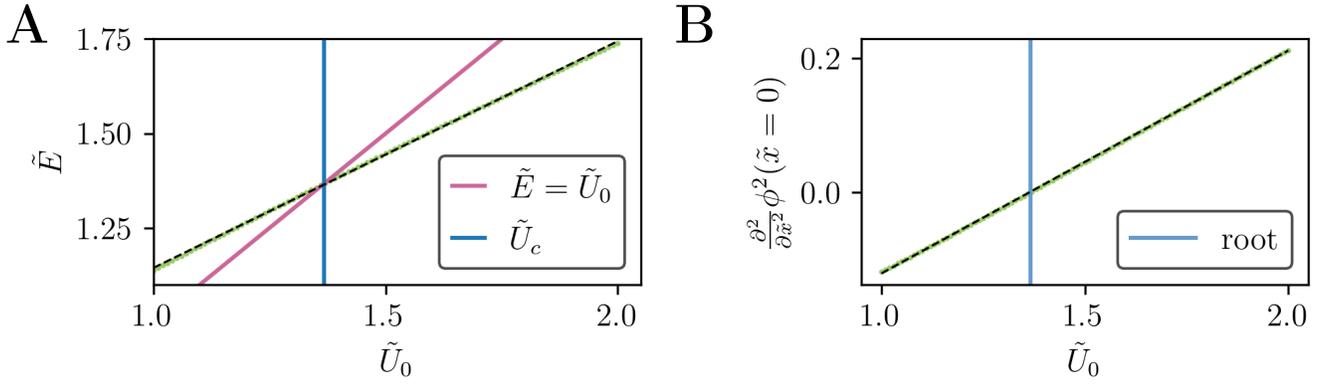

  \centering
  \begin{overpic}[width=.49\textwidth]{{/../figs/groundstate_energy_over_barrier_alt_symbol2}.png}
  \put(-2,56){\huge \bf A}
\end{overpic}
  \begin{overpic}[width=.49\textwidth]{{/../figs/groundstate_curv_over_barrier}.png}
  \put(-2,56){\huge \bf B}
\end{overpic}
  \caption{{\bf A} Dimensionless ground state energy $\widetilde E$ as a function of the
   barrier height $ \widetilde U_0$ (green dots).
  A linear fit  according to $ \widetilde E(U_0) = 0.601\, \widetilde U_0 + 0.542$  is shown as a black dashed line.
  The crossing between the ground-state energy $\widetilde E$ and $\widetilde E=\widetilde U_0$ (red line) is marked as a vertical blue line at $\widetilde U_c := \widetilde U_0(\widetilde E = \widetilde U_0) = 1.36$.
  {\bf B} Curvature of the ground state probability density at the origin as a function of  $ \widetilde U_0$.
  A linear fit according to  $\partial^2 \phi^2/{\partial \tilde x}^2 (\tilde x=0) = 0.334\,  \widetilde  U_0-0.455$ is shown as
  a black dashed line. The point where the curvature vanishes  is marked as a vertical blue line,
   which is given by  $ \widetilde U_c = 1.36$.}
  \label{fig:energy_over_barrier}
\end{figure}

\begin{figure}
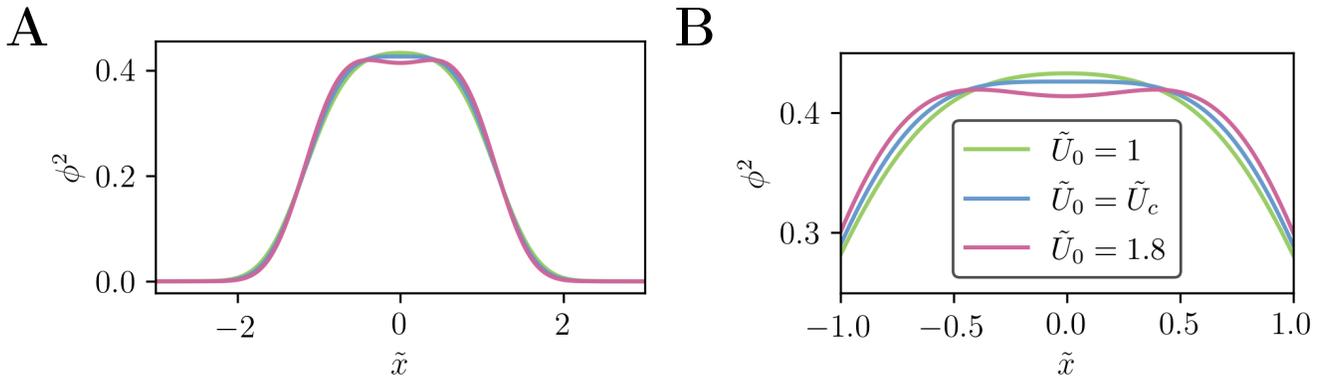

  \centering
\begin{overpic}[width=.49\textwidth]{{/../figs/groundstate_wf}.png}
 \put(-2,56){\huge \bf A}
\end{overpic}
\begin{overpic}[width=.49\textwidth]{{/../figs/groundstate_wf_zoom}.png}
 \put(-2,56){\huge \bf B}
\end{overpic}
  \caption{{\bf A} Ground state probability densities for  $\widetilde U_0=1$, $\widetilde U_0= \widetilde U_c$ and $\widetilde U_0=1.8$. {\bf B} Ground state probability densities around the origin.}
  \label{fig:density-compare}
\end{figure}

Dimensionless ground state energies $\widetilde E$ are  numerically
computed for $\widetilde U_0\in[1,2]$ using bisection to  find the eigenvalues of the dimensionless Schr\"odinger equation eq.~\eqref{eqn:SE}. The ground state energy $\widetilde E$ shows approximately linear dependence on $\widetilde U_0\in[1,2]$ (see green dots in fig.~\ref{fig:energy_over_barrier}A). A  minimum of the ground-state density
is defined by a positive curvature at the origin, $\frac{\partial^2}{{\partial x}^2}\phi^2 (\tilde x=0)>0$. From the Schr\"odinger equation eq.~\eqref{eqn:SE} it becomes clear that the wave function of the ground state develops a minimum at the origin for $\widetilde E<\widetilde U_0$. It follows for the curvature of the probability density
\begin{align}
  \frac{\partial^2}{{\partial \tilde x}^2} \phi^2 &=  2\left[ \left(\frac{\partial \phi}{\partial \tilde x}\right)^2 + (\widetilde U(\tilde x) - \widetilde E)\phi^2\right].
\end{align}
Since the ground state is symmetric around the origin this becomes at the origin
\begin{equation}
  \frac{\partial^2}{{\partial \tilde x}^2} \phi^2 = 2(\widetilde U_0 - \widetilde E)\phi^2.
\end{equation}
Fig.~\ref{fig:energy_over_barrier}B shows the curvature at the origin $\frac{\partial^2}{{\partial x}^2} \phi ^2 (\tilde x=0)$
as a function of  $\widetilde U_0$ (green dots). A linear fit is used to determine the critical potential strength at which
the curvature vanishes as
$\widetilde U_c = 1.36$. Fig.~\ref{fig:density-compare} shows a comparison of the ground state density
distributions  for  $\widetilde U_0=\{1, \widetilde U_c, 1.8\}$.

\begin{figure}[b]
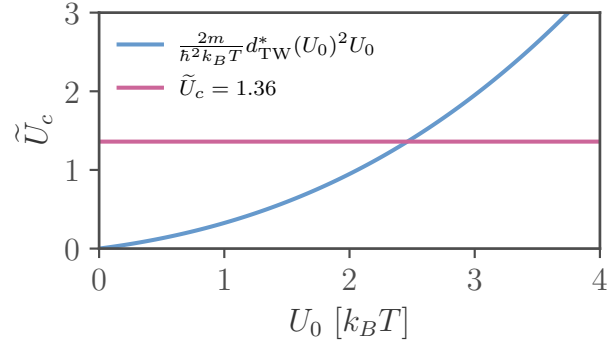

  \centering
\begin{overpic}[width=.49\textwidth]{{/../figs3/fe/zundel_UcoverU}.pdf}
 \put(-2,56){}
\end{overpic}
  \caption{A minimum in the ground state probability density appears for $\frac{2m }{\hbar^2 k_BT} {{d_{\rm TW}^*}^2(U_0)} U_0 > \widetilde U_c$ as a rearrangement of eq.~\eqref{eq:barrierCondition} with $U_0$ being the barrier height of the double-well potential, $\widetilde U_c = 1.36$ taken from fig.~\ref{fig:energy_over_barrier}B and ${d_{\rm TW}^*(U_0)}$ taken from the linear fit in fig.~\ref{dwFits}B.}
\label{effBarrier}
\end{figure}

From this analysis we conclude that when treating the excess proton quantum-mechanically,
a minimum in the ground state probability density, reflecting the effect of the barrier, appears for barrier heights of the double-well potential
\begin{align}
 U_0 > \frac{\hbar^2}{2m {{d_{\rm TW}^*}^2(U_0)}}\,\widetilde U_c\,k_BT,
 \label{eq:barrierCondition}
 %= 2.5\,k_BT.
\end{align}
which holds for $U_0>2.5\,k_BT$ as can be read off fig.~\ref{effBarrier}.
 Therefore, using the naive assumption that a minimum in the probability density at the barrier top indicates the presence of an effective quantum-mechanical barrier, which is far from obvious, one could speculate that the presented results for the spectroscopic signatures of proton barrier-crossing are expected to survive quantum-mechanical zero-point motion effects for high enough potential barriers.

\clearpage

\section{Summary of previous experimental and theoretical studies on  infrared spectra of the H$_5$O$_2{}^{+}$ cation}
\label{pubDataSection}

\begin{figure*}[htp]
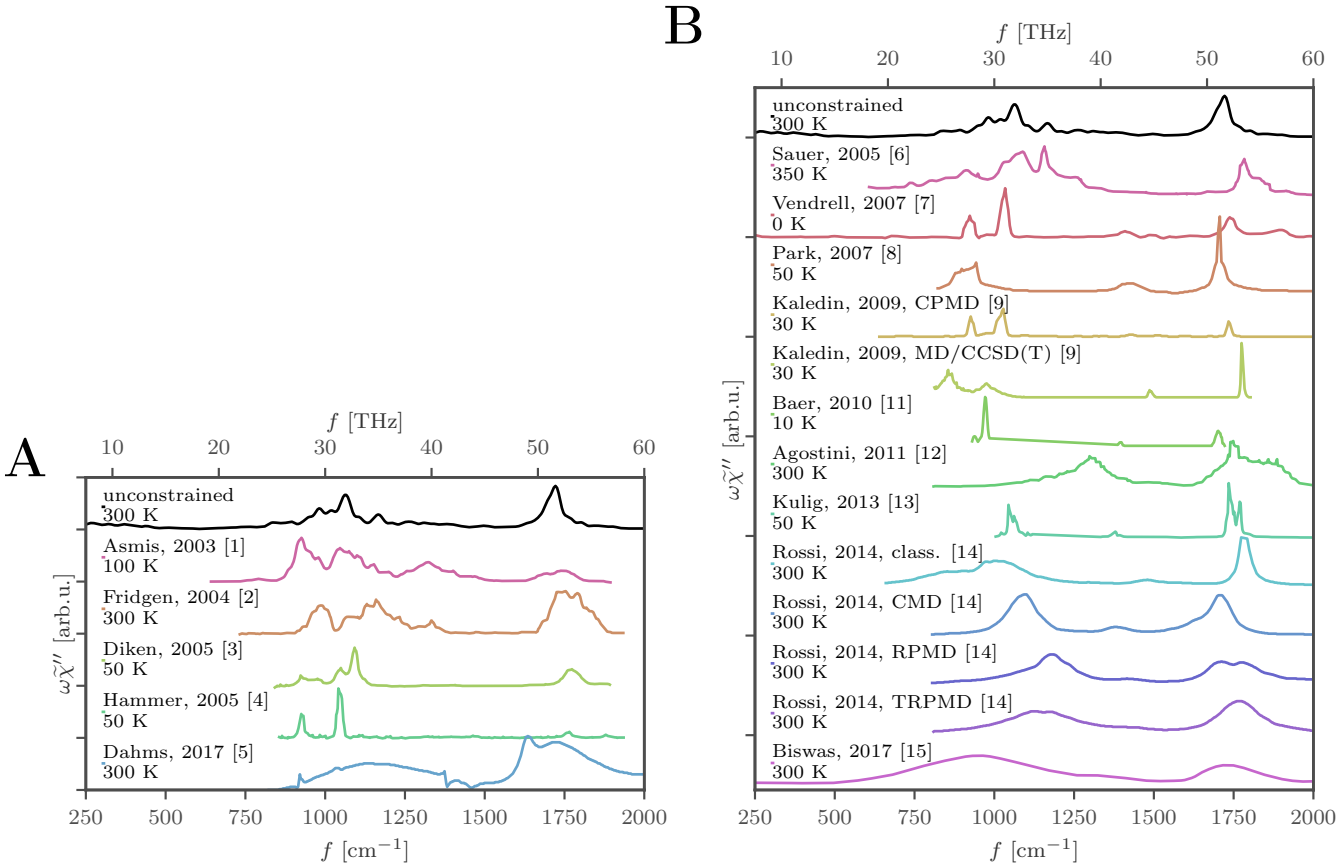

\centering
\begin{overpic}[width=0.49\textwidth]{{/../figs/zundel_compare_exp}.pdf}
\put(-4,62){\huge \bf A}
\end{overpic}
\begin{overpic}[width=0.49\textwidth]{{/../figs/zundel_compare_theo}.pdf}
\put(-4,95){\huge \bf B}
\end{overpic}
\caption{Collection of experimental (A) and theoretical (B) \ac{IR} spectra of the H$_5$O$_2{}^{+}$ cation in the proton-transfer regime (colored lines), compared to the \ac{IR} spectrum obtained from \ac{AIMD} simulations in this study, shown in black. {\bf A} The experimental spectra were recorded using multiple-photon-dissociation (IRPMD) spectroscopy \cite{Asmis2003a, Fridgen2004}, predissociation (IRPD) spectroscopy in Argon \cite{Diken2005} and Neon \cite{Hammer2005} and Fourier-transform \ac{IR} (FTIR) spectroscopy of H$_5$O$_2{}^{+}$ cations solvated in acetonitrile \cite{Dahms2017}. {\bf B} Theoretical spectra were obtained using \ac{AIMD} simulations on the MP2/cc-pVTZ level \cite{Sauer2005a}, the multiconfiguration time-dependent Hartree (MCTDH) method \cite{Vendrell2007a}, CPMD simulations using the BLYP functional \cite{Park2007}, CPMD simulations using the BLYP functional \cite{Kaledin2009}, \ac{MD} simulations on the CCSD(T) potential energy surface using MP2 dipole moment functions (based on \cite{Huang2005}) \cite{Kaledin2009}, \ac{AIMD} simulations on the BLYP-D3 TZV2P level \cite{Baer2010}, \ac{MD} simulations using the multistate-empirical-valence-bond (MS-EVB) method \cite{Agostini2011}, \ac{AIMD} simulations on the DZVP-BLYP level \cite{Kulig2013}, classical \ac{MD}, centroid \ac{MD} (CMD), as well as ring polymer \ac{MD} simulations (RPMD, TRPMD)\cite{Rossi2014} on the CCSD(T) potential energy surface \cite{Huang2005} and normal mode analysis on the 6-311++G(d,p)-B3LYP level performed on states obtained from a multistate-empirical-valence-bond (MS-EVB) \ac{MD} simulation \cite{Biswas2017}.
All theoretical results were obtained for a single H$_5$O$_2{}^{+}$ cation, except for \cite{Biswas2017}, who computed the spectra from clusters of 16--18 water molecules and an excess-proton.
}
 \label{zundel_compare_exp_theo}
\end{figure*}

\begin{figure*}
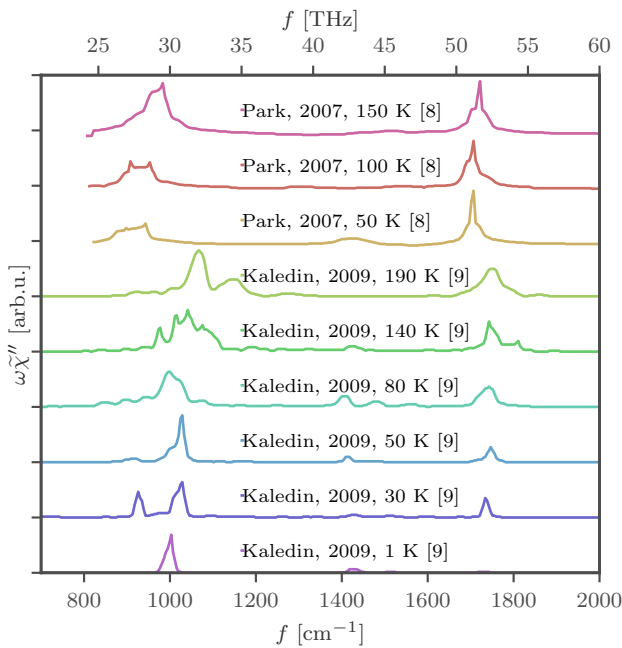

\begin{overpic}[width=0.49\textwidth]{{/../figs/zundel_compare_theo_T}.pdf}
\end{overpic}
\caption{Comparison of theoretical \ac{IR} spectra of the H$_5$O$_2{}^{+}$ cation in the proton-transfer regime at various temperatures obtained from Car-Parinello \ac{MD} (CPMD) simulations using the BLYP density functional method \cite{Park2007,Kaledin2009}.}
 \label{zundel_compare_theo_T}
\end{figure*}

A collection of  \ac{IR} spectra of the H$_5$O$_2{}^{+}$ cation in the so-called proton-transfer regime, \SIrange{600}{1500}{cm^{-1}}, obtained from experiments is given in fig.~\ref{zundel_compare_exp_theo}A and from theoretical calculations in fig.~\ref{zundel_compare_exp_theo}B. The spectra are compared to the spectrum of the unconstrained H$_5$O$_2{}^{+}$  cation obtained from \ac{AIMD} simulations in this study (black lines on top). The spectra vary greatly among different experimental studies in this regime, highlighting the   subtle differences between experimental techniques and also pointing to a pronounced
 temperature dependence of the spectra. \citet{Fridgen2004} say that ``at this point, we are unable to determine the source of the discrepancy between the present infrared spectrum and that obtained by \citet{Asmis2003a}''. \citet{Sauer2005a} discuss ``that differences between the IRMPD spectrum \cite{Asmis2003a} and the IRPD spectrum \cite{Diken2005} should be due to the different excitation mechanisms and/or different temperatures''. \citet{Hammer2005} state ``the \ac{IR} profiles obtained in these two measurements \cite{Asmis2003a, Fridgen2004} were markedly different, perhaps reflecting the different ion sources used in the two experiments and/or the specific fluence characteristics of the laser sources''. \citet{Vendrell2007a} introduce the topic stating that ``spectra could not be consistently assigned in terms of fundamental frequencies and overtones of harmonic vibrational modes due to large amplitude anharmonic displacements and couplings of the cluster \cite{Asmis2003a, Fridgen2004, Headrick2004, Hammer2005}'' and conclude their study by pointing out that their ``reported calculations are in excellent agreement to the experimental measurements of Refs. \cite{Headrick2004, Hammer2005} on the predissociation spectrum of H5O2+·Ne.'', which is the most widely accepted low-temperature spectrum to date. Furthermore, nuclear quantum effects were studied in detail by \citet{Rossi2014}, who compared spectra from different ring polymer \ac{MD} simulations, and \citet{Park2007} and \citet{Baer2010} successfully investigated the messenger-induced changes of  spectra apparent in IRMPD techniques. 
The temperature dependence was analyzed by \citet{Park2007} and \citet{Kaledin2009} in Car-Parinello \ac{MD} (CPMD) simulations, presented in fig.~\ref{zundel_compare_theo_T}, and classical \ac{MD} simulations on the CCSD(T) potential energy surface using MP2 dipole moment functions (based on \cite{Huang2005}), stating that ``classical MD simulations at a temperature of \SI{30}{K} qualitatively reproduce many of the key features in the experimental vibrational [Ar-] predissociation''. More recent studies have focused on the H$_5$O$_2{}^{+}$ cation in larger water clusters or in bulk water or other solvents and at room temperature \cite{Thaemer2015, Kulig2013, Dahms2017, Biswas2017, Agostini2011, Fournier2018}. These studies identify remarkably broad \ac{IR} features in the \SIrange{1000}{1200}{cm^{-1}} regime, associated with ``proton shuttling'' \cite{Thaemer2015}, the ``proton-transfer mode (PTM)'' \cite{Kulig2013,Biswas2017, Dahms2017} or the ``shared proton stretch'' \cite{Fournier2018}.

The review of the published experimental and theoretical data shows that the data produced in this study is in qualitative agreement with previous results. However, especially in the region of \SIrange{800}{1200}{cm^{-1}} associated with the proton-transfer motion, large temperature and system-dependent differences between the previously published results are observed and reflect
 the incomplete understanding of proton-transfer dynamics.

\newpage

\section{Normal-mode analysis of the H$_5$O$_2{}^{+}$ cation}
\label{normalModeSection}

\begin{figure*}[tbh]
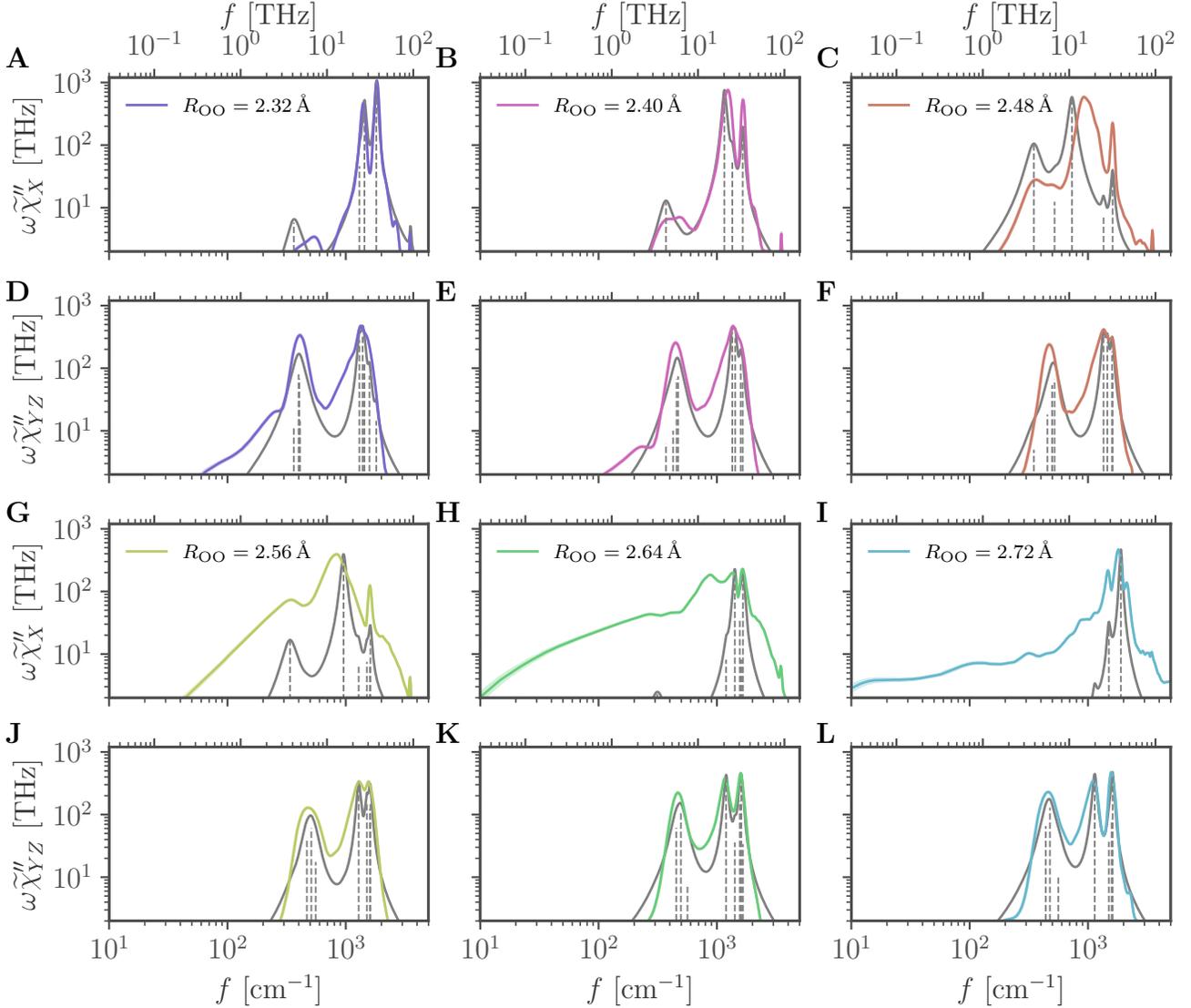

\centering
\begin{overpic}[width=\textwidth]{{/../figs/zundel_compare_spectra_nm_xyz_all}.pdf}
\put(2,79){\large \bf A}
\put(37,79){\large \bf B}
\put(68,79){\large \bf C}
\put(2,60){\large \bf D}
\put(37,60){\large \bf E}
\put(68,60){\large \bf F}
\put(2,42){\large \bf G}
\put(37,42){\large \bf H}
\put(68,42){\large \bf I}
\put(2,24){\large \bf J}
\put(37,24){\large \bf K}
\put(68,24){\large \bf L}
\end{overpic}
\caption{\ac{IR} spectra of the excess proton (colored lines) in the H$_5$O$_2{}^{+}$ cation for different constraint oxygen distances $R_{\mathrm{OO}}$ compared to normal-mode spectra calculated from energetically optimized structures and projected on the proton coordinate(grey dashed sticks). The discrete normal modes sum to smooth spectra (grey lines) by including line-broadening at finite temperature obtained from damped harmonic oscillations of the normal modes according to eq.~\ref{LinResHO} in section \ref{hoResponseSection} and assuming a friction coefficient of $\gamma=$\SI{16}{u/ps}. The spectra are shown for different orientations ({\bf A--C, G--I}: $x$-axis connecting the two oxygens, {\bf D--F, J--L}: $yz$-plane).}
\label{allNormalModes}
\end{figure*}

The normal modes are obtained as the Eigenvectors of the Hessian of the optimal structure in a chosen coordinate system, the Eigenvalues give the frequencies. The normal-mode analysis of the H$_5$O$_2{}^{+}$ systems was performed for energetically optimized structures using the same set of parameters as for the \ac{AIMD} simulation.

A projection of the Eigenvectors on the proton coordinate allows to obtain the relative magnitudes of the excess-proton spectra predicted from the normal-mode analysis. These are scaled to the magnitudes of the excess-proton spectra obtained directly from the \ac{AIMD} simulations for various $R_{\mathrm{OO}}$ and orientations in fig.~\ref{allNormalModes}. To improve the normal-mode spectra, line-broadening of the normal modes at finite temperature is modeled by damped harmonic oscillations (see SI section \ref{hoResponseSection}).
In the $yz$-plane the agreement is very good for all barrier heights indicating, that the method is well suited for modeling the excess-proton dynamics without a barrier. Along $x$ the normal mode spectra serve to explain the major peaks of the spectra at high frequencies which will be identified as the normal-mode contributions. The broad low-frequency shoulder, which is related to the barrier-crossing effects, can obviously not be modeled using normal modes.

The system with $R_{\mathrm{OO}}=\SI{2.40}{\ang}$ represents the global unconstrained minimum of the free-energy landscape, shown in fig.~\ref{zundel_intro}D in the main text, and can therefore be used to compare qualitatively to published data on the unconstrained H$_5$O$_2{}^{+}$ cation, as done in tab.~\ref{tab:nm} for the range \SIrange{100}{2000}{cm^{-1}}. The normal-modes can be grouped to rocking and wagging modes in the range \SIrange{300}{600}{cm^{-1}}, the water-water stretch at around \SIrange{550}{650}{cm^{-1}}, which is missing in the constrained case, the widely discussed proton-transfer mode along $x$, which varies greatly between \SIrange{800}{1200}{cm^{-1}}, a pair of modes between \SIrange{1300}{1600}{cm^{-1}}, associated with perpendicular excess-proton motion in the $yz$-plane, and the two water-bending modes at \SIrange{1600}{1800}{cm^{-1}}, for which the lower one shows excess-proton motion in the $yz$-plane (named in-phase or gerade) and the higher one shows excess-proton motion along $x$ (named out-of-phase or ungerade).

\begin{figure*}[p]
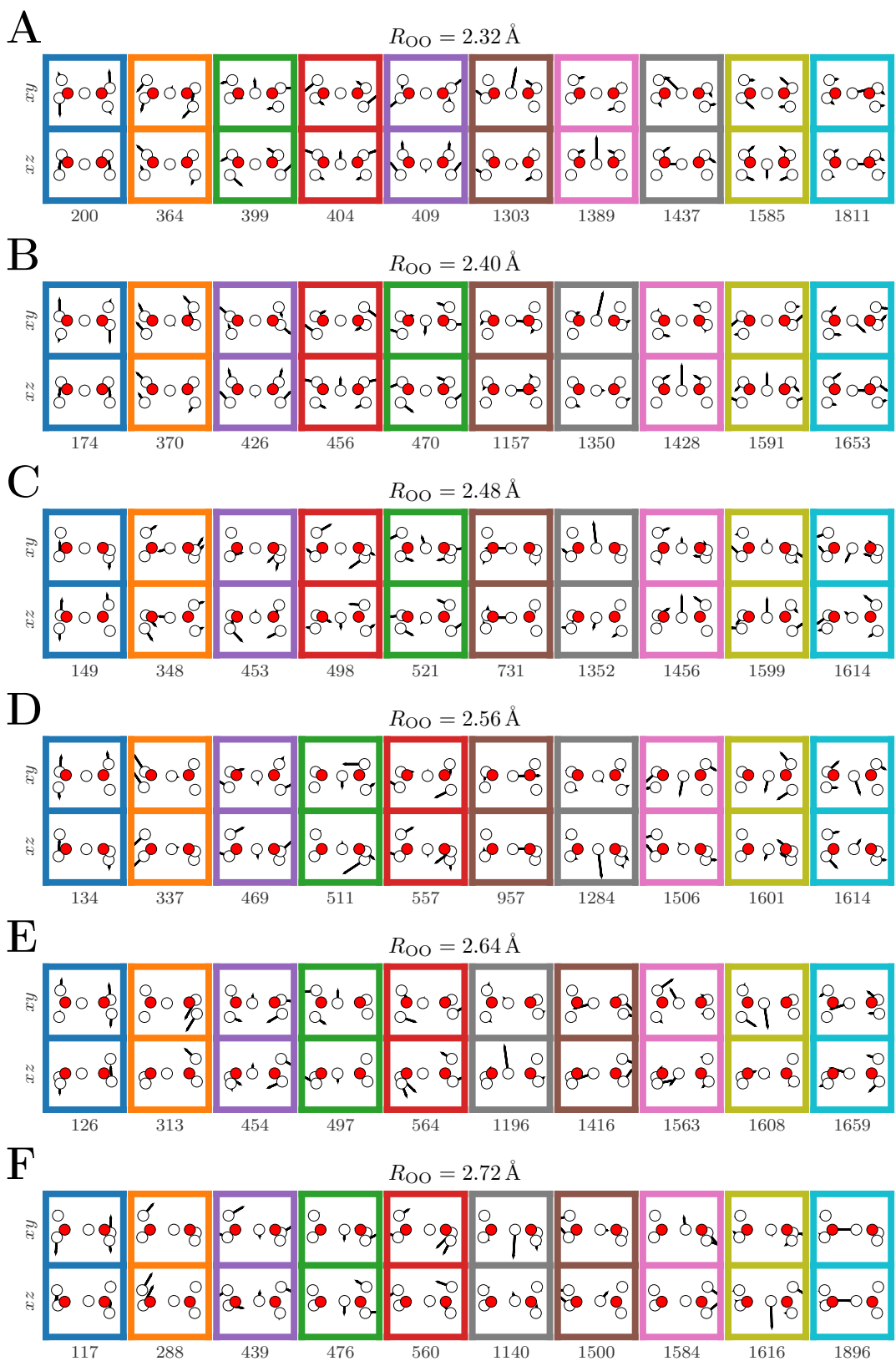

\centering
\begin{overpic}[width=0.8\textwidth]{{/../figs3/nm/zundel_nm_vis_d2.32}.pdf}
\put(0,22){\huge \bf A}
\end{overpic}
\begin{overpic}[width=0.8\textwidth]{{/../figs3/nm/zundel_nm_vis_d2.40}.pdf}
\put(0,22){\huge \bf B}
\end{overpic}
\begin{overpic}[width=0.8\textwidth]{{/../figs3/nm/zundel_nm_vis_d2.48}.pdf}
\put(0,22){\huge \bf C}
\end{overpic}
\begin{overpic}[width=0.8\textwidth]{{/../figs3/nm/zundel_nm_vis_d2.56}.pdf}
\put(0,22){\huge \bf D}
\end{overpic}
\begin{overpic}[width=0.8\textwidth]{{/../figs3/nm/zundel_nm_vis_d2.64}.pdf}
\put(0,22){\huge \bf E}
\end{overpic}
\begin{overpic}[width=0.8\textwidth]{{/../figs3/nm/zundel_nm_vis_d2.72}.pdf}
\put(0,22){\huge \bf F}
\end{overpic}
\caption{Illustrations of the normal modes of the H$_5$O$_2{}^{+}$ cation for various constrained $R_{\mathrm{OO}}$, projected on the $xy$ and $xz$ planes. The colored frames guide the eye through the shifting of a respective normal mode through A-F.}
\label{NormalModesVis}
\end{figure*}

\begin{figure*}[bth]
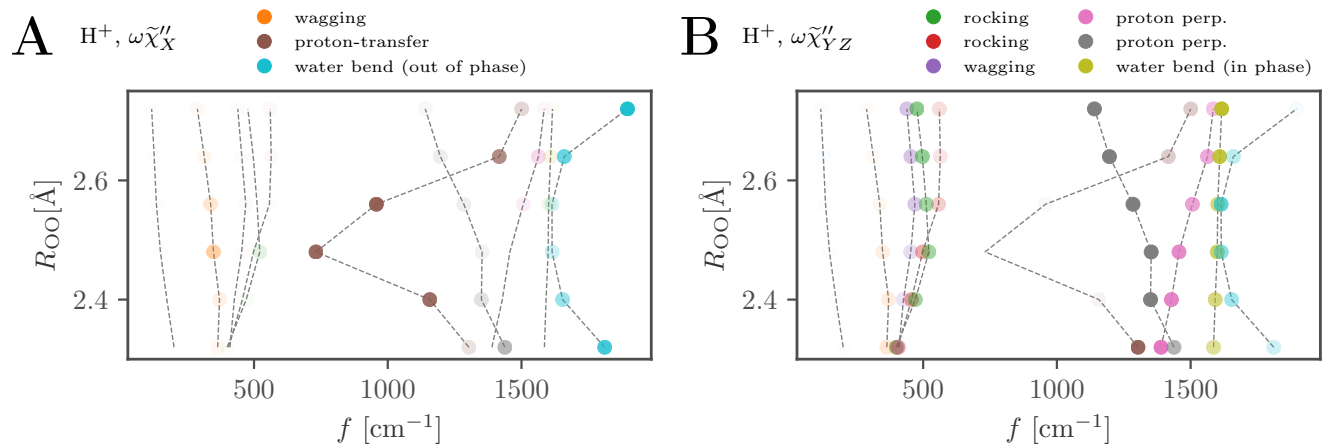

\centering
\begin{overpic}[width=0.49\textwidth]{{/../figs3/nm/zundel_nm_allD_2_x}.pdf}
\put(0,62){\huge \bf A}
\end{overpic}
\begin{overpic}[width=0.49\textwidth]{{/../figs3/nm/zundel_nm_allD_2_yz}.pdf}
\put(0,62){\huge \bf B}
\end{overpic}
\caption{Normal mode frequencies of the excess-proton dynamics in the H$_5$O$_2{}^{+}$ cation for various constrained $R_{\mathrm{OO}}$ as shown in fig.~\ref{allNormalModes}. The shading indicates the relative amplitudes of the respective normal modes, normalized to the maximal amplitude for each value of $R_{\mathrm{OO}}$. The colors and grey dashed lines guide the eye through the shifting of a respective normal mode, as also shown in fig.~\ref{NormalModesVis}.}
\label{compareNormalModes}
\end{figure*}

Illustrations of the normal modes in the range \SIrange{100}{2000}{cm^{-1}} of H$_5$O$_2{}^{+}$ cations with various contrained $R_{\mathrm{OO}}$ are shown in fig.~\ref{NormalModesVis}, where the colored frames indicate similar normal modes for different values of $R_{\mathrm{OO}}$. Fig.~\ref{compareNormalModes} summarizes the frequencies of the normal modes for all values of $R_{\mathrm{OO}}$ and highlights the associated relative excess-proton-motion intensities either along $x$ or in the $yz$-plane. Most modes shift only weakly, including the wagging and rocking modes in the range \SIrange{300}{600}{cm^{-1}} and the pair of modes between \SIrange{1300}{1600}{cm^{-1}}, associated with perpendicular excess-proton motion in the $yz$-plane. However, the proton-transfer mode ($\SI{1157}{cm^{-1}}$ for $R_{\mathrm{OO}}=\SI{2.40}{\ang}$) strongly increases for values of $R_{\mathrm{OO}}\geq \SI{2.40}{\ang}$, as shown previously by \citet{Wolke2016} for the D$_5$O$_2{}^{+}$ cation, which possibly explains why the proton-transfer mode in bulk is suggested to reside at much higher wavenumbers compared to the gas-phase spectra \cite{Kulig2013}. The in-phase water bending mode shows no shifting but the out-of-phase water bending mode, associated with excess-proton motion along $x$, strongly shifts. Generally, modes associated with proton motion along $x$ are shown to be highly sensitive on the value of $R_{\mathrm{OO}}$, while modes that are associated with proton motion in the $yz$-plane are not. %The splitting in the OH-stretching regime at around \SI{3500}{cm^{-1}} for increased $R_{\mathrm{OO}}$ is well visible as well as slowing down of the slowest modes.

\begin{turnpage}
\begin{table}[hp]
\centering
\tiny
\caption{Collection of previous identifications of normal modes in the H$_5$O$_2{}^{+}$ cation in the range \SIrange{100}{2000}{cm^{-1}}, compared to data obtained in this study with constrained $R_{\mathrm{OO}}=\SI{2.40}{\ang}$ in the first row. Bold fonts indicate significant motion of the excess proton (if reported).}
\label{tab:nm}
\vspace{10pt}
\begin{tabular}{c | c c  c c c c  c c  c c }
BLYP-D3, TZV2P, $R_{\mathrm{OO}}=\SI{2.40}{\ang}$ & wagging, rocking  & 320, 426, \bf{456}, \bf{470} & & & p. trans. ($x$) & \bf{1157} & proton perp. (${yz}$) & \bf{1350}, \bf{1428} & water bend & \bf{1591}, \bf{1653}  \\ \hline
MP2, aug-cc-pVTZ \cite{Sauer2005a} & H$_2$O wag, H$_2$O rock & 362, 461, 532, 535 & sym. OHO str.  & 627 & asym. OHO str. & 884 & OHO bend $(y,z)$ & 1484, 1557  & HOH bend & 1710, 1765\\
MP2, cc-pVTZ \cite{Sauer2005a} & & 367, 451, 522, 529 & & 631  & & 961 & & 1493, 1572 & (in/out phase) & 1710, 1773   \\
MP2, cc-pVTZ(aug-O) \cite{Sauer2005a} & & 347, 456, 525, 532, & & 626  & & 904 & & 1473, 1551 &  &   1706, 1761\\
CCSD(T), cc-pVTZ \cite{Sauer2005a} & & 350, 457, 531, 535 & & 633  & & 896 & & 1493, 1572 & & 1724, 1780 \\ \hline
CCSD(T), MP2, aug-cc-pVTZ \cite{Huang2005} &  & 339, 471, 532, 554 &  & 630 &  & 861 &  & 1494, 1574 & & 1720, 1770 \\ \hline
MP2, aug-cc-pVTZ \cite{Hammer2005} & & & & &  OHO stretch  & 808   & & & HOH bend & 1662, 1717 \\ \hline
MCTDH \cite{Vendrell2007a}         & wagging & 106, 108, 232, 254, 374, 422  & wat-wat str.      & 550 &  p. trans. doub. & \bf{918}, \bf{1033} & proton perp. & \bf{1391} & water bend & 1606, 1741 \\
                                   & rocking &  481, 915, 930,  943 & & &  &  & p. trans. + wat. str. & \bf{1411}  & (gerade, unger.)  &  \\ \hline
BLYP, aug-cc-pVTZ \cite{Park2007} & & & & & asym. OHO str. & 938 & OHO bend (y, z) & 1408, 1478 & HOH bend & 1647, 1702 \\
B3LYP, aug-cc-pVTZ \cite{Park2007} & & &  & & & 891 & & 1397, 1455 & & 1619, 1678 \\
MP2, aug-cc-pVTZ \cite{Park2007} & & &  & & & 868 & & 1405, 1478 & & 1627, 1680 \\ \hline
BLYP-D3, TZV2P, AIMD \cite{Baer2010} &  wagging, rocking & 370, 431, 498, 514 & OO stretch & 571 & p. trans. ($v_q$) & \bf{971} & proton$_{yz}$ ($\gamma_p, \gamma_p'$) & \bf{1396}, 1477 &  water bend ($\delta_s, \delta_a$) & 1636, \bf{1700} \\ \hline
MS-EVB, EMA \cite{Agostini2011} & & & & & & 1032 &  & 1311, 1560 &   & 1638, 1769 \\ \hline
B3LYP, aug-cc-pVTZ \cite{Kulig2013} & & 358 & & & & 926 &  &  1446 & & 1742\\
\end{tabular}
\end{table}
\end{turnpage}

\clearpage

\section{Decomposition of the H$_5$O$_2{}^{+}$ \ac{IR} spectra into excess-proton and water contributions}
\label{hDecompSection}

\begin{figure*}[pth]
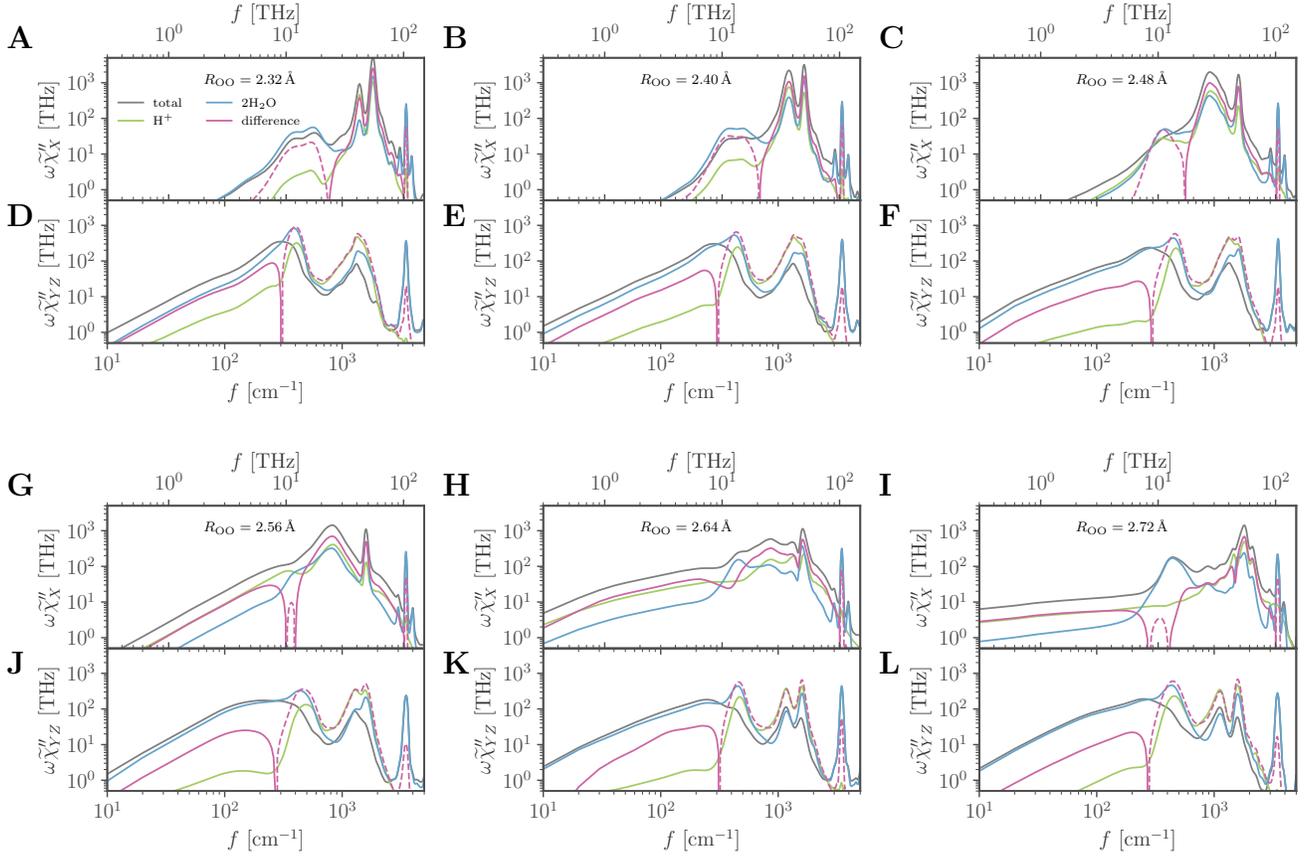

\centering
\begin{overpic}[width=0.32\textwidth]{{/../figs3/decomp/zundel_decomp_d2.32_f6_loglog}.pdf}
\put(-2,85){\large \bf A}
\put(-2,45){\large \bf D}
\end{overpic}
\begin{overpic}[width=0.32\textwidth]{{/../figs3/decomp/zundel_decomp_d2.40_f6_loglog}.pdf}
\put(-2,85){\large \bf B}
\put(-2,45){\large \bf E}
\end{overpic}
\begin{overpic}[width=0.32\textwidth]{{/../figs3/decomp/zundel_decomp_d2.48_f6_loglog}.pdf}
\put(-2,85){\large \bf C}
\put(-2,45){\large \bf F}
\end{overpic}
\begin{overpic}[width=0.32\textwidth]{{/../figs3/decomp/zundel_decomp_d2.56_f6_loglog}.pdf}
\put(-2,85){\large \bf G}
\put(-2,45){\large \bf J}
\end{overpic}
\begin{overpic}[width=0.32\textwidth]{{/../figs3/decomp/zundel_decomp_d2.64_f6_loglog}.pdf}
\put(-2,85){\large \bf H}
\put(-2,45){\large \bf K}
\end{overpic}
\begin{overpic}[width=0.32\textwidth]{{/../figs3/decomp/zundel_decomp_d2.72_f6_loglog}.pdf}
\put(-2,85){\large \bf I}
\put(-2,45){\large \bf L}
\end{overpic}
\caption{Decomposition of the \ac{IR} spectra of the H$_5$O$_2{}^{+}$ cation with two constrained oxygens at a given distance $R_{\mathrm{OO}}$ in different directions ({\bf A--C, G--I}: along the $x$-axis connecting the two oxygens, {\bf D--F, J--L}: along the $yz$-plane) into water and excess-proton contributions. The total \ac{IR} spectra are shown in grey, the spectra of the excess proton in green and the spectra of the two water molecules in blue. The cross-correlation spectrum defined as $\widetilde \chi''_{\mathrm{H}^+,\mathrm{H}_2\mathrm{O}}=\widetilde \chi''_{\mathrm{tot}}- \widetilde \chi''_{\mathrm{H}^+}- \widetilde \chi''_{\mathrm{H}_2\mathrm{O}}$ is shown in red. The dashed red line denotes negative values of the cross-correlation spectrum.
}
\label{allDecompSpectra}
\end{figure*}

By using Wannier centers for charge localization, the total dipole moment of the simulation systems can be exactly decomposed into proton and water contributions $\mathbf{p}_{\mathrm{tot}}(t)=\mathbf{p}_{\mathrm{H}^+}(t)+\mathbf{p}_{\mathrm{H}_2\mathrm{O}}(t)$. A comparison of the \ac{IR} spectra of the total dipole moment $\omega \widetilde \chi''_{\mathrm{tot}}$ of the H$_5$O$_2{}^{+}$ cation for various constrained oxygen positions to the power spectra of only the excess proton $\omega \widetilde \chi''_{\mathrm{H}^+}$ and the power spectra of the dipole moments of the two flanking water molecules $\omega \widetilde \chi''_{\mathrm{H}_2\mathrm{O}}$ is shown in fig.~\ref{allDecompSpectra}A-L in different directions. As discussed in SI section~\ref{linearResponseSection}, the cross-correlation spectra $\omega \widetilde \chi''_{\mathrm{H}^+,\mathrm{H}_2\mathrm{O}}$, shown in red in fig.~\ref{allDecompSpectra}A-L, are proportional to the cross-correlations of water-dipole-moment and excess-proton dynamics. Along $x$, the cross-correlation spectra of the six systems shown here, are nearly entirely positive as well as nearly proportional to the power spectrum of the proton itself, indicating constructive coupling of the proton motion to the water dipole moments along this axis, as previously shown \cite{Sauer2005a}. An apparent exception to this is the rocking and wagging regime at \SIrange{300}{600}{cm^{-1}} and the split OH-stretching mode at around \SI{3500}{cm^{-1}}, which shows a strong negative cross-correlation spectrum in both $x$ direction and $yz$-plane. Adjacent to this band all systems show two weak OH-stretching vibrations along $x$, likely a Stark effect of the mean excess-proton field on the water dipole moments.

In the $yz$-plane the cross-correlation spectra are nearly entirely negative, indicating mainly out-of-phase motion of the excess-proton and the water dipole moments, producing generally weaker spectra in $yz$ compared to $x$.

It can be concluded that the \ac{IR} spectrum of the H$_5$O$_2{}^{+}$ cation along the axis connecting the two water oxygens reflects the excess-proton dynamics. This justifies the \ac{IR} signal to be used as a reporter of the excess-proton dynamics, as well as the focus on its dynamics.

\newpage

\section{Comparison of constrained and unconstrained dichroic \ac{IR} spectra of the H$_5$O$_2{}^{+}$ cation}
\label{dichroicSpectraSection}

\begin{figure*}[!hp]
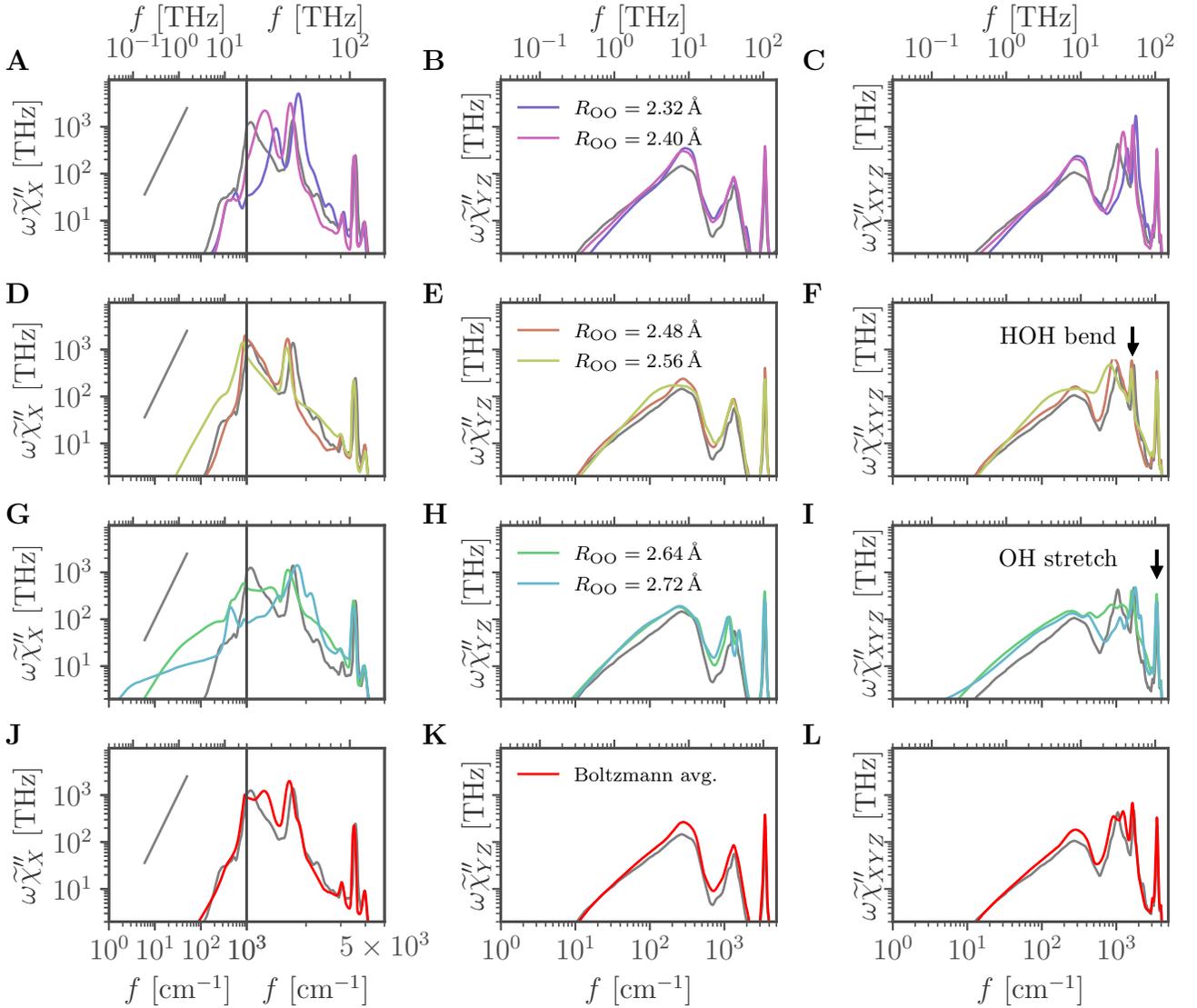

\centering
\begin{overpic}[width=\textwidth]{{/../figs/zundel_compare_spectra_Boltzmann_xyz_all_x_2}.pdf}
\put(2,79){\large \bf A}
\put(36,79){\large \bf B}
\put(67,79){\large \bf C}
\put(2,60){\large \bf D}
\put(36,60){\large \bf E}
\put(67,60){\large \bf F}
\put(2,42){\large \bf G}
\put(36,42){\large \bf H}
\put(67,42){\large \bf I}
\put(2,24){\large \bf J}
\put(36,24){\large \bf K}
\put(67,24){\large \bf L}
\end{overpic}
\caption{\ac{IR} spectra of the H$_5$O$_2{}^{+}$ cation in different directions ({\bf A, D, G, J}: along the $x$-axis connecting the two oxygens, {\bf B, E, H, K}: along the $yz$-plane {\bf C, F, I, L}: isotropic spectrum). {\bf A--I} The colors correspond to systems with different fixed oxygen-oxygen separation $R_{\mathrm{OO}}$, grey lines denote the unconstrained system. {\bf J--L} Comparison of the \ac{IR} spectra of the unconstrained system (grey lines) to a Boltzmann average (red lines) according to eq.~\eqref{eq:Boltzmann_spectra} of the spectra in A-I. The HOH-bending mode and OH-stretching mode of the water molecules are indicated in F and I. The short straight grey lines indicate the $\omega^2$-scaling associated with the low-frequency shoulder of the power spectra.}
\label{zundel_compare_spectra_xyz}
\end{figure*}

\ac{IR} spectra of the unconstrained H$_5$O$_2{}^{+}$ cation and various systems with constrained $R_{\mathrm{OO}}$ are shown in fig.~\ref{zundel_compare_spectra_xyz} in different directions. The $x$-axis corresponds to the direction connecting the two oxygens, while the $yz$-plane is orthogonal to that axis, as illustrated in fig.~\ref{zundel_intro}A in the main text.
Interestingly, the spectra in $x$-direction strongly depend on the value of the constrained oxygen separation $R_{\mathrm{OO}}$ for frequencies lower than \SI{2500}{cm^{-1}} (figs.~\ref{zundel_compare_spectra_xyz}A, \ref{zundel_compare_spectra_xyz}D, \ref{zundel_compare_spectra_xyz}G and \ref{zundel_compare_spectra_xyz}J). The HOH-bending mode is shifted to almost \SI{2000}{cm^{-1}} for $R_{\mathrm{OO}}=\SI{2.32}{\ang}$ (fig.~\ref{zundel_compare_spectra_xyz}A) and to slightly lower frequencies than the unconstrained system for $R_{\mathrm{OO}}\geq\SI{2.40}{\ang}$ (figs.~\ref{zundel_compare_spectra_xyz}D and \ref{zundel_compare_spectra_xyz}G), whereas the frequency of the OH-stretching mode is not affected by fixing $R_{\mathrm{OO}}$.

The various spectra for the $yz$-plane in figs.~\ref{zundel_compare_spectra_xyz}B, \ref{zundel_compare_spectra_xyz}E, \ref{zundel_compare_spectra_xyz}H and \ref{zundel_compare_spectra_xyz}K are indistinguishable on the other hand. The dominant features are the librations of the water molecules at \SI{300}{cm^{-1}} and the OH-stretching mode at around \SI{3500}{cm^{-1}}. While the OH-stretching mode appears in both directions, the HOH-bending mode is mainly visible along the $x$-axis. Note that in the $yz$-plane the HOH-bending mode contributions of the outer hydrogens of each water molecule cancel out due to symmetric motion with respect to the $x$-axis.

The isotropic spectra in figs.~\ref{zundel_compare_spectra_xyz}C, \ref{zundel_compare_spectra_xyz}F, \ref{zundel_compare_spectra_xyz}I and \ref{zundel_compare_spectra_xyz}L depend on $R_{\mathrm{OO}}$ only in the regime \SIrange{400}{1700}{cm^{-1}} since for lower frequencies, between \SIrange{10}{400}{cm^{-1}}, the $yz$-contributions to the isotropic spectrum are dominant. In particular, this motivates the analysis of dynamics along $x$-direction, which shows the dominant contribution to the isotropic spectrum except for very low frequencies.

Interestingly, the spectra of the unconstrained systems are well recovered by a Boltzmann average of the spectra of the systems with fixed $R_{\mathrm{OO}}$ according to
\begin{align}
\label{eq:Boltzmann_spectra}
\omega \widetilde \chi''(\omega)_{<R_{\mathrm{OO}}>} = \frac{\sum_{i} \omega \widetilde \chi''(\omega)_{R_{\mathrm{OO}_i}} e^{-U(R_{\mathrm{OO}_i})/k_BT}}{\sum_{i} e^{-U(R_{\mathrm{OO}_i})/k_BT}},
\end{align}
using the free energy along $R_{\mathrm{OO}}$ of the unconstrained system, $U(R_{\mathrm{OO}})$, shown in fig.~\ref{zundel_intro}D in the main text. As shown in figs.~\ref{zundel_compare_spectra_xyz}J, fig.~\ref{zundel_compare_spectra_xyz}K and fig.~\ref{zundel_compare_spectra_xyz}L, the agreement is very good along all directions, which indicates a sufficient dynamic decoupling of the slow oxygen coordinate $R_{\mathrm{OO}}$ from the proton coordinate $d$ with respect to \ac{IR} spectra and allows observations for the constrained systems to be generalized to the unconstrained system.

\section{Recrossing transfer paths of the excess-proton in the H$_5$O$_2{}^{+}$ cation}
\label{recrossingSection}

\begin{figure}[bth]
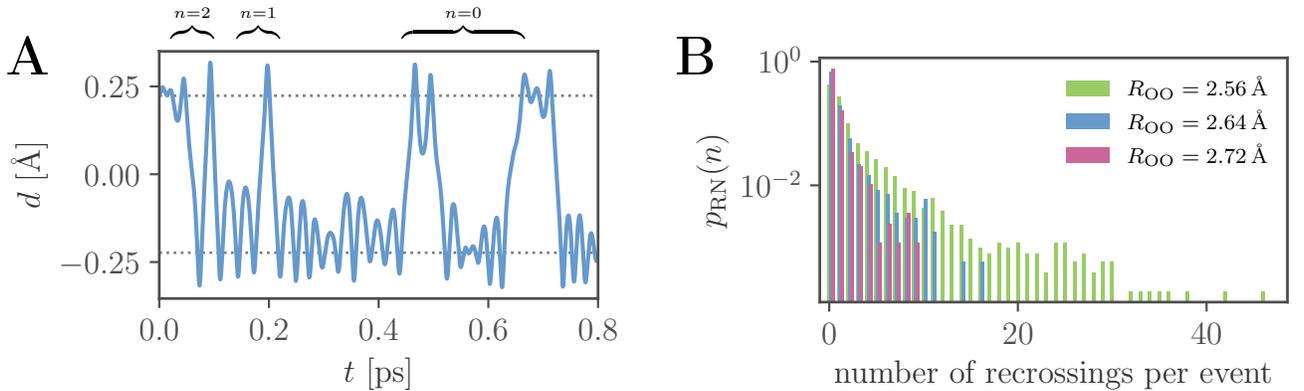

\centering
\begin{overpic}[width=0.49\textwidth]{{/../figs/zundel_traj_recrExp}.pdf}
\put(2,53){\huge \bf A}
\put(27,58){$ \overbrace{\quad}^{n=2}$}
\put(37,58){$ \overbrace{\ }^{n=1}$}
\put(62,58){$ \overbrace{\quad \quad \quad \quad \quad}^{n=0}$}
\end{overpic}
\begin{overpic}[width=0.49\textwidth]{{/../figs3/tp/zundel_tp_recrStat}.pdf}
\put(2,53){\huge \bf B}
\end{overpic}
\caption{\textbf{A} Example trajectory of the excess-proton coordinate $d(t)$ in the H$_5$O$_2{}^{+}$ cation with constrained $R_{\mathrm{OO}}$=\SI{2.64}{\ang}, showing nine transfer paths between the minima of the free energy denoted as grey dotted lines. The first three transfer paths belong to a single transfer event (with two recrossings). The subsequent two transfer paths also belong to a single transfer event (with one recrossing). The remaining four transfer paths are all defined to not be recrossing. It follows $p_{\mathrm{RN}}(0)=4/6$, $p_{\mathrm{RN}}(1)=1/6$ and $p_{\mathrm{RN}}(2)=1/6$. \textbf{B} Recrossing-number probability distributions $p_{\mathrm{RN}}(n)$ normalized such that $\sum_{n=0}^{\infty}p_{\mathrm{RN}}(n)=1$, for different $R_{\mathrm{OO}}$.}
\label{recrStat}
\end{figure}

After crossing the barrier once, the excess proton often immediately recrosses the barrier. In order to quantify this effect, the transfer paths are grouped into transfer events. A transfer event is defined by subsequently occurring transfer paths without recrossing of the same minimum and is sorted by the number of these crossings. An example for this definition is given in fig.~\ref{recrStat}A. The normalized distribution of the transfer events by the number of recrossings, in the following called the recrossing-number distribution $p_{\mathrm{RN}}(n)$, is given in fig.~\ref{recrStat}B. For low barriers up to 40 subsequent recrossings are observed. For higher barriers the distribution is shifted to lower numbers of recrossings. Nevertheless, for all barrier heights a significant fraction of transfer events consist of multiple recrossings, $\sum_{i=1}^{\infty}p_{\mathrm{RN}}(n)=0.47$ for $R_{\mathrm{OO}}=\SI{2.56}{\ang}$, 0.32 for $R_{\mathrm{OO}}=\SI{2.64}{\ang}$ and 0.14 for $R_{\mathrm{OO}}=\SI{2.72}{\ang}$. The mean number of recrossings is $\sum_{i=0}^{\infty}n\ p_{\mathrm{RN}}(n)=1.84$ for $R_{\mathrm{OO}}=\SI{2.56}{\ang}$, 0.70 for $R_{\mathrm{OO}}=\SI{2.64}{\ang}$ and 0.41 for $R_{\mathrm{OO}}=\SI{2.72}{\ang}$.

\section{\ac{IR} power spectra from linear-response theory}
\label{linearResponseSection}

Assuming linear response of an observable $x(t)$ with respect to a force that couples to an observable $y(t)$, the response function $\chi_{xy}(t)$ is related to the correlation function $C_{xy}(t')=\langle x (t+t') y(t)\rangle$
\begin{align}
\label{linearResponseSI}
\chi_{xy}(t) = - \frac{1}{ k_BT } \Theta(t)\frac{d}{dt} C_{xy}(t),
\end{align}
where $k_BT$ is the thermal energy. Realizing that $\chi(t)$ is single-sided, the Fourier transform is calculated as
\begin{align}
\widetilde \chi_{xy}(\omega) &= - \frac{1}{ k_BT } \int_{0}^{\infty} dt\ e^{i\omega t} \frac{d}{dt} C_{xy}(t)\nonumber \\
 &= - \frac{1}{ k_BT } \left( C_{xy}(0) - i \omega \int_0^{\infty} dt\ e^{i\omega t} C_{xy}(t) \right)\nonumber \\
 &= - \frac{1}{ k_BT } \left( C_{xy}(0) - i \omega \widetilde C_{xy}^+(\omega) \right),
\label{linearResponseFTSI}
\end{align}
where the superscript $^+$ denotes a single-sided Fourier transform.
In case of $x=y$, $C_{xx}(t)$ is an autocorrelation function, which is real and symmetric, therefore it follows for the imaginary part of the response function in Fourier space
\begin{align}
\widetilde \chi_{xx}''(\omega) &= \frac{1}{ k_BT } \omega \operatorname{Re} ( \widetilde  C^+_{xx}(\omega) ) \\
\label{omegaDoublePrimeCXX}
 &= \frac{1}{ k_BT } \frac{\omega}{2} \widetilde C_{xx}(\omega).
\end{align}
When computing the power spectra of a stochastic process $x(t)$, limited to the time domain $[0,L_t]$, the Wiener-Kintchine theoreme, eq.~\eqref{WienerKintchineFT} in section \ref{WienerKintchineSection}, can be used to express $\widetilde C_{xx}(\omega)$ in terms of $\tilde x(\omega)$, turning eq.~\eqref{omegaDoublePrimeCXX} into
\begin{align}
\label{omegaDoublePrimeXOmega}
\widetilde \chi_{xx}''(\omega) =
\frac{\omega}{2 k_BT L_t } | \tilde x(\omega)|^2.
\end{align}

In case of $x(t)$ being the polarization $\bm p(t)$ of the system, which is coupled to an external electric field $\bm E(t)$, the dimensionless dielectric susceptibility $\chi(t)$ is given by
\begin{align}
\widetilde \chi(\omega) =  \frac{1}{V\epsilon_0 l} \langle \widetilde \chi_{\bm p \bm p}(\omega) \rangle,
\end{align}
where $\epsilon_0$ is the vacuum permittivity, $V$ is the system volume and an average is performed over the $l$ dimensions of $\bm p$.

\section{Spectral cross contributions of excess-proton dynamics in the H$_5$O$_2{}^{+}$ cation}
\label{crossContribSection}

A decomposition of a trajectory $x(t)$ into two parts $x (t)=x_1 (t)+ x_2 (t)$ gives rise to three contributions in the total power spectrum
\begin{align}
\nonumber
\omega \widetilde \chi_{xx}''(\omega)
&= \frac{\omega^2}{2 k_BT} \left[ \widetilde C_1(\omega)+ \widetilde C_2(\omega)+ 2\widetilde C_{1,2}(\omega) \right]\\
&= \omega \left[\widetilde \chi''_1(\omega) + \widetilde \chi''_2(\omega) + \widetilde \chi''_{1,2}(\omega) \right],
\end{align}
where the cross-correlation contribution $\chi''_{1,2}(\omega)$ is defined such that it equals the difference spectrum
\begin{align}
\nonumber
\widetilde \chi''_\mathrm{diff}(\omega)&= \widetilde \chi_{xx}''-\widetilde \chi''_1(\omega) - \widetilde \chi''_2(\omega) = \widetilde \chi''_{1,2}(\omega) \\
&= \frac{\omega}{k_BT} \widetilde  C_{1,2}(\omega).
\end{align}
A positive cross-correlation spectrum hints to in-phase motion, a negative cross-correlation spectrum to out-of-phase motion of $x_1 (t+t')$ and $x_2 (t)$ at a given frequency.

\begin{figure*}[tbh]
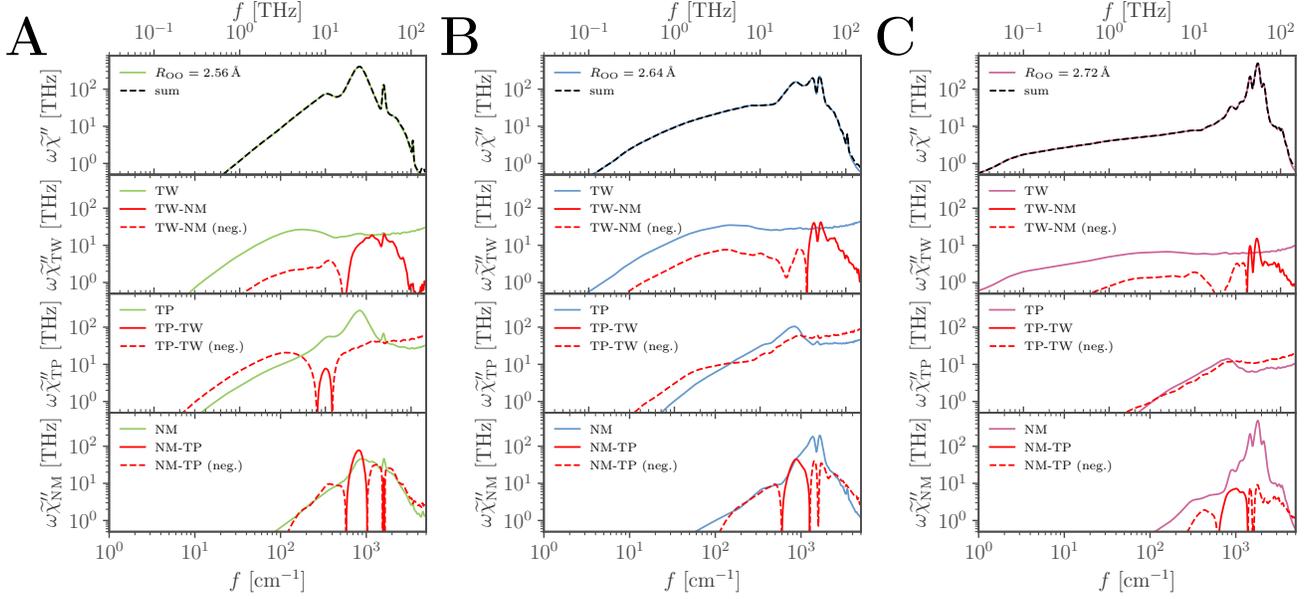

\centering
\begin{overpic}[width=0.32\textwidth]{{/../figs3/tp/zundel_tp_decompCrossD_d2.56_log}.pdf}
\put(-2,88){\huge \bf A}
\end{overpic}
\begin{overpic}[width=0.32\textwidth]{{/../figs3/tp/zundel_tp_decompCrossD_d2.64_log}.pdf}
\put(-2,88){\huge \bf B}
\end{overpic}
\begin{overpic}[width=0.32\textwidth]{{/../figs3/tp/zundel_tp_decompCrossD_d2.72_log}.pdf}
\put(-2,88){\huge \bf C}
\end{overpic}
\caption{Spectral decomposition of the excess-proton dynamics in the H$_5$O$_2{}^{+}$ cation for different constrained $R_{\mathrm{OO}}$ ({\bf A}: $R_{\mathrm{OO}}=\SI{2.56}{\ang}$, {\bf B}: $R_{\mathrm{OO}}=\SI{2.64}{\ang}$, {\bf C}: $R_{\mathrm{OO}}=\SI{2.72}{\ang}$), similar  to the results in fig.~\ref{decomp}B in the main text (colored lines). The different cross-contributions are shown in red and negative cross-contributions as red broken lines.}
\label{crossContribSI}
\end{figure*}

Therefore the decomposition of the excess-proton dynamics in the time domain according to $d(t)=d_{\mathrm{TW}}(t)+d_{\mathrm{TP}}(t)+d_{\mathrm{NM}}(t)$ into a transfer-waiting $d_{\mathrm{TW}}(t)$, a transfer-path $d_{\mathrm{TP}}(t)$ and a normal-mode contribution $d_{\mathrm{NM}}(t)$ as introduced in fig.~\ref{decomp}, in the main text, produces spectral cross-contributions. The spectral contributions $\omega \widetilde \chi''_{\mathrm{TW}}$, $\omega \widetilde \chi''_{\mathrm{TP}}$ and $\omega \widetilde \chi''_{\mathrm{NM}}$ as well as the spectral cross contributions $\omega \widetilde \chi''_{\mathrm{TW,NM}}$, $\omega \widetilde \chi''_{\mathrm{TP,TW}}$ and $\omega \widetilde \chi''_{\mathrm{NM,TP}}$ resulting from this decomposition are shown in fig.~\ref{crossContribSI}.

\newpage

\section{Dynamics of a two-state process}
\label{binaryResponseSection}

\begin{figure}[tb]
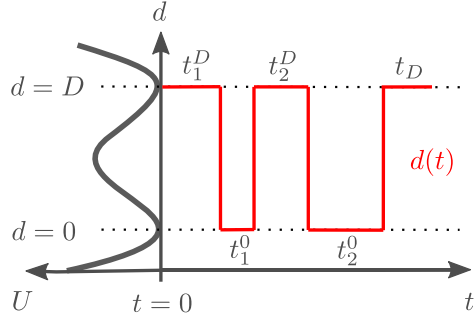

\begin{overpic}[width=0.35\textwidth]{{/../figs/barrier_theory4}.pdf}
\end{overpic}
\caption{Schematic of a binary jump process $d(t)$ with jumps separated by residence times $t_j^0$ and $t_j^D$ following the notation considered in eq.\eqref{pLL}. A doublewell potential, U($d$), is shown to highlight the relation of the binary jump process to the barrier-crossing dynamics.}
\label{binary_jump}
\end{figure}

Consider an arbitrary two-state process $d(t)$ characterized by a a jump distance $D$, a jump-time probability distribution, which is the transfer-waiting-time probability distribution $p_{\mathrm{TW}}(t)$ with $\int_0^{\infty} p_{\mathrm{TW}}(t) dt = 1$, and a survival probability $q_{\mathrm{TW}}(t) = \int_t^{\infty} p_{\mathrm{TW}}(t') dt'$. The conditional expectation values $\langle d(t) \rangle |_{d(0)}$ for the process starting at time $t=0$ at either $d(0)=0$ or $d(0)=D$ are thus given as
\begin{align}
\langle d(t) \rangle |_{d(0)=0} &= \cancel{0\ p_{00}(t)} + D\ p_{0D}(t),\\
\langle d(t) \rangle |_{d(0)=D} &= \cancel{0\ p_{D0}(t)} + D\ p_{\mathrm{DD}}(t),
\end{align}
where $p_{xy}(t)$ denotes the probability to be at $d(t)=y$ when starting at $d(0)=x$. In the long-time limit each state has equal probability. The autocorrelation function, $C(t) =  \langle d(t) d(0) \rangle$, is therefore given as
\begin{align}
\langle d(t) d(0) \rangle &= \frac{1}{2} \left(\cancel{0\ \langle d(t) \rangle |_{d(0)=0}} +D\ \langle d(t) \rangle |_{d(0)=D} \right) \\
&= \frac{D^2}{2} p_{\mathrm{DD}}(t).
\label{}
\end{align}
In order to express the probability $p_{\mathrm{DD}}(t)$ in terms of the first-passage-time distribution $p_{\mathrm{TW}}(t)$ and survival probability $q_{\mathrm{TW}}(t)$, all possible jumps within time $t$ have to be considered, as illustrated in fig.~\ref{binary_jump}
\begin{align}
\begin{split}
p_{\mathrm{DD}}(t) = &\int_0^{\infty} dt_D q_{\mathrm{TW}}(t_D) \sum_{N=0}^{\infty} \prod_{j=1}^N \int_0^{\infty} dt_j^D p_{\mathrm{TW}}(t_j^D )\\
&\int_0^{\infty} dt_j^0 p_{\mathrm{TW}}(t_j^0 )\ \delta \left[t-t_D-\sum_{j=1}^N (t_j^D+t_j^0)\right].
\end{split}
\label{pLL}
\end{align}

Since $p_{\mathrm{TW}}(t)$ and $q_{\mathrm{TW}}(t)$ are single-sided, a single-sided Fourier transform is performed,
\begin{align}
\begin{split}
\tilde p_{\mathrm{DD}}^+(\omega) = &\int_0^{\infty} dt\ e^{i\omega t} \int_0^{\infty} dt_D q_{\mathrm{TW}}(t_D) \sum_{N=0}^{\infty} \prod_{j=1}^N \int_0^{\infty} dt_j^D p_{\mathrm{TW}}(t_j^D ) \\
&\int_0^{\infty} dt_j^0 p_{\mathrm{TW}}(t_j^0 )\ \delta \left[t-t_D-\sum_{j=1}^N (t_j^D+t_j^0)\right]
\end{split}\\
\begin{split}
 = &\int_0^{\infty} dt_D q_{\mathrm{TW}}(t_D) \sum_{N=0}^{\infty} \prod_{j=1}^N \int_0^{\infty} dt_j^D p_{\mathrm{TW}}(t_j^D ) \\
&\int_0^{\infty} dt_j^0 p_{\mathrm{TW}}(t_j^0 )\ e^{i\omega  \left[t_D+\sum_{j=1}^N (t_j^D+t_j^0)\right]}
\end{split}\\
\label{pLLFT}
 = &\tilde q_{\mathrm{TW}}(\omega) \sum_{N=0}^{\infty} \tilde p_{\mathrm{TW}}(\omega)^{2N} \\
 = &\frac{\tilde q_{\mathrm{TW}}(\omega)}{1 - \tilde p_{\mathrm{TW}}(\omega)^2},
\end{align}
and it follows for the single-sided Fourier transformed autocorrelation
\begin{align}
\widetilde C^+(\omega) = \frac{D^2}{2} \frac{ \tilde q_{\mathrm{TW}}(\omega)}{1 - \tilde p_{\mathrm{TW}}(\omega)^2 }.
\label{jumpAutocorrelationSI}
\end{align}

From eqs.~\eqref{linearResponseFTSI} and \eqref{jumpAutocorrelationSI} we obtain our expression for the response function of the binary jump process

\begin{align}
\widetilde \chi_{DD}(\omega) = -\frac{D^2}{2 k_BT}  \left(1- \frac{ i\omega \tilde q_{\mathrm{TW}}(\omega)}{1 - \tilde p_{\mathrm{TW}}(\omega)^2}\right).
\end{align}

In case of the jumping variable $d(t)$ being a polarization with polarization jump $2qd^*_{\rm TW}$, where $q$ is the charge, the dielectric susceptibility of the binary polarization jump process reads

\begin{align}
\widetilde \chi(\omega) = -\frac{2q^2{d^*_{\rm TW}}^2}{ V \epsilon_0 k_BT}  \left(1- \frac{ i\omega \tilde q_{\mathrm{TW}}(\omega)}{1 - \tilde p_{\mathrm{TW}}(\omega)^2}\right),
\end{align}
and eventually its power spectrum, proportional to the imaginary part of the dielectric susceptibility, is obtained as
\begin{align}
\omega \widetilde \chi''(\omega) = \frac{2q^2{d^*_{\rm TW}}^2}{ V \epsilon_0 k_BT}\ \text{Re}\left(\frac{\omega^2 \tilde q_{\mathrm{TW}}(\omega)}{1 - \tilde p_{\mathrm{TW}}(\omega)^2}\right).
\end{align}

\newpage

\section{Power spectrum of the damped harmonic oscillator}
\label{hoResponseSection}

The absorbed power $\omega \widetilde \chi_{xx}''(\omega)$ of the damped harmonic oscillator described by the differential equation
\begin{align}
m\ddot x(t) = -\gamma \dot x(t) - k x(t) + F_{\mathrm{ext}} (t),
\label{HO}
\end{align}
is computed from the linear response in Fourier space
\begin{align}
\widetilde \chi_{xx}(\omega) &= \frac{\tilde x(\omega)}{\tilde F_{\mathrm{ext}}(\omega)} \\
&= (k-m\omega^2-i\gamma\omega)^{-1} \\
&= \frac{k-m\omega^2+i\gamma\omega}{(k-m\omega^2)^2+\gamma^2\omega^2},
\end{align}
where $\tilde x(\omega)$ is the oscillating variable, $m$ the mass, $\gamma$ the friction coefficient, $k$ the spring constant of the harmonic potential and $\tilde F_{\mathrm{ext}}(\omega)$ an external force.
For the power spectrum follows
\begin{align}
\omega \widetilde \chi_{xx}''(\omega)  = \frac{\gamma\omega^2}{(k-m\omega^2)^2+\gamma^2\omega^2},
\end{align}
which by introducing the time scales $\tau = 2 \gamma / k$, $\tau_m = \sqrt{m / k}$ and length scale $D$ with $D^2 = k_BT/k$ converts to
\begin{align}
\omega \widetilde \chi_{xx}''(\omega)  = \frac{2D^2}{k_BT}\frac{\tau \omega^2}{4(1-\tau_m^2 \omega^2)^2+\tau^2\omega^2}.
\end{align}
In case of the oscillating variable $x(t)$ being a polarization with polarization jump $qD$, where $q$ is the charge, the dielectric susceptibility reads
\begin{align}
\omega \widetilde \chi''(\omega)  = \frac{2q^2D^2}{V\epsilon_0k_BT}\frac{\tau \omega^2}{4(1-\tau_m^2 \omega^2)^2+\tau^2\omega^2}.
\label{LinResHO}
\end{align}

In spectroscopy this is known as a Lorentz band shape, which in the overdamped case, $\tau_m \to 0$, reads
\begin{align}
\omega \widetilde \chi''(\omega)  = \frac{2q^2D^2}{V\epsilon_0k_BT}\frac{\tau \omega^2}{4+\tau^2\omega^2},
\label{OverdampedDebyeSpectrum}
\end{align}
known as the Debye band shape \cite{debye1929polar}.

\newpage
\section{Transfer-path shape}
\label{tpPathIntegralSection}

\citet{Kim2015} derived the transfer-path-time shape $t(d_\text{TP})$ over a harmonic barrier by an exact calculation, valid for arbitrary barrier height, as
\begin{align}
t(d_\text{TP})&= \nonumber\\
&{\tau}-\frac{2 \gamma {d^*_{\mathrm{TW}}}^2}{ k_BT U}\int_{\sqrt{U}}^{\sqrt{U}(d_\text{TP}/d^*_{\mathrm{TW}}-1)}\text{d}y\left( \frac{\text{erf}(y) - \text{erf}(\sqrt{U})}{\text{erf}(\sqrt{U}(d_\text{TP}/d^*_{\mathrm{TW}}-1)) - \text{erf}(\sqrt{U})} - \frac{1}{2}\right)D_{+}(y)
, \label{eq:harm_shape1}
\end{align}
where $D_{+}(x)=e^{-x^2}\int_{0}^{x}dt\ e^{t^2}$ is the Dawson integral function, $\gamma$ is the friction constant, $U=U_0/k_BT$ is the dimensionless barrier height, $d^*_{\mathrm{TW}}$ a length scale and $\tau$ a time scale. Note that the shape function is given by time $t$ as a function of position $d$. The second term in eq.~\eqref{eq:harm_shape1} vanishes for
 $d_\text{TP} = 2 d^*_{\mathrm{TW}}$ and reduces to $-{\tau}$ for  $d_\text{TP} = 0$ and therefore $t(d_\text{TP}=0)=0$ and $t(d_\text{TP}=2 d^*_{\mathrm{TW}})={\tau}$.
For a variable $s=y/\sqrt{U}$, eq.~\eqref{eq:harm_shape1} is rewritten as
\begin{align}
t(d_\text{TP})&= \nonumber\\
&{\tau}-\frac{2 \gamma {d^*_{\mathrm{TW}}}^2}{ k_BT \sqrt{U}}\int_{1}^{d_\text{TP}/d^*_{\mathrm{TW}}-1}\text{ds}\left( \frac{\text{erf}(\sqrt{U}s) - \text{erf}(\sqrt{U})}{\text{erf}(\sqrt{U}(d_\text{TP}/d^*_{\mathrm{TW}}-1)) - \text{erf}(\sqrt{U})} - \frac{1}{2}\right)D_{+}(\sqrt{U}s)
, \label{eq:harm_shape2}
\end{align}

For large $U \gg 1$ and $d^*_{\mathrm{TW}}<d_\text{TP}<2 d^*_{\mathrm{TW}}$, eq.~\eqref{eq:harm_shape2} has an asymptotic expression
\begin{align}
t(d_\text{TP})
&\approx
{\tau} + \frac{\gamma {d^*_{\mathrm{TW}}}^2}{k_BT }\int_{1}^{d_\text{TP}/d^*_{\mathrm{TW}}-1}\text{d}t \frac{D_{+}(\sqrt{U}t)}{\sqrt{U}} \nonumber\\
&= {\tau} + \frac{\gamma {d^*_{\mathrm{TW}}}^2}{2 k_BT U} \ln (d_\text{TP}/d^*_{\mathrm{TW}}-1)
, \label{eq:harm_shape3}
\end{align}
where we use $D_{+}(\sqrt{U}t) / \sqrt{U} \approx 1/(2Us)$.

Therefore, further using the symmetric nature of $t(d_\text{TP})$ in the limit $U \rightarrow \infty$, we obtain the asymptotic expression for $t(d_\text{TP})$ as

\begin{eqnarray}
t(d_\text{TP})
=
\begin{cases}
- \frac{\gamma {d^*_{\mathrm{TW}}}^2}{2 k_BT U}\ln (1 - d_\text{TP}/d^*_{\mathrm{TW}}),~&\text{for}~0<d_\text{TP}<d^*_{\mathrm{TW}} \\
{\tau} + \frac{\gamma {d^*_{\mathrm{TW}}}^2}{2 k_BT U} \ln (d_\text{TP}/d^*_{\mathrm{TW}}-1),~&\text{for}~d^*_{\mathrm{TW}}<d_\text{TP}<2 d^*_{\mathrm{TW}}.
\end{cases}
\label{eq:harm_shap4}
\end{eqnarray}
It is straightforward to solve eq.~\eqref{eq:harm_shap4} for $d_\text{TP}$, yielding
\begin{eqnarray}
d_\text{TP}(t)
=
\begin{cases}
-d^*_{\mathrm{TW}}  \left(e^{-\frac{2 k_BT U t}{\gamma {d^*_{\mathrm{TW}}}^2}}-1\right),~&\text{for}~0<t<{\tau}/2 \\
d^*_{\mathrm{TW}}  \left(e^{\frac{2 k_BT U (t-\tau)}{\gamma {d^*_{\mathrm{TW}}}^2}}+1\right),~&\text{for}~{\tau}/2<t<{\tau}.
\end{cases}
\label{eq:harm_shap5}
\end{eqnarray}
Using the curvature parameter $\kappa = \gamma {d^*_{\mathrm{TW}}}^2/(2 k_BT U)$, and shifting the variables $t \rightarrow t + {\tau}/2$ and $d_\text{TP} \rightarrow d_\text{TP} + d^*_{\mathrm{TW}}$ to fulfill $d_\text{TP}(t=0)=0$, we arrive at the leading order expression as the sum of the above two
\begin{eqnarray}
d_\text{TP}(t)
&=
d^*_{\mathrm{TW}} \left[ e^{-\frac{\tau -2 t}{2 \kappa }} -  e^{-\frac{\tau +2 t}{2 \kappa }} \right] \nonumber \\
&= d^*_{\mathrm{TW}} \left[ e^{ t/ \kappa } -  e^{- t/ \kappa } \right]/ e^{\frac{ \tau}{ 2 \kappa }} .
\label{eq:harm_shap6}
\end{eqnarray}

This expression is easily compared to the transfer-path shape (in the presence of a harmonic potential) derived from the path-integral approach (equivalent to eq.~3 in the main text) \cite{Faccioli2006, Cossio2018}
\begin{align}
 d_\text{TP}(t)={d^*_{\mathrm{TW}}}
 \left[ e^{ t / \kappa} - e^{-t / \kappa} \right]
 /{\cal N},
 \label{eq:tp_shape_eq3}
\end{align}
with a slightly different normalization factor ${\cal N} = e^{\frac{\tau}{2\kappa}}-e^{-\frac{\tau}{2\kappa}}$. Note that the difference vanishes in the high-barrier limit $U \rightarrow \infty$, i.e. $\kappa \rightarrow 0$, in which limit eq.~\eqref{eq:harm_shap6} derived from reference \cite{Kim2015} becomes equivalent to eq.~\eqref{eq:tp_shape_eq3} derived from the path-integral approach.

\newpage
\section{Spectral signatures of transfer paths}
\label{theoryTPSection}

The \ac{IR} spectral signature of transfer paths is derived by modeling the Fourier-transformed transfer-path contribution $\tilde d_{\mathrm{TP}}(\omega)$ based on the transfer-path time $\tau_{\mathrm{TP}}$ and the recrossing-number probability distribution $p_{\mathrm{RN}}(n)$.

\begin{figure}[hb]
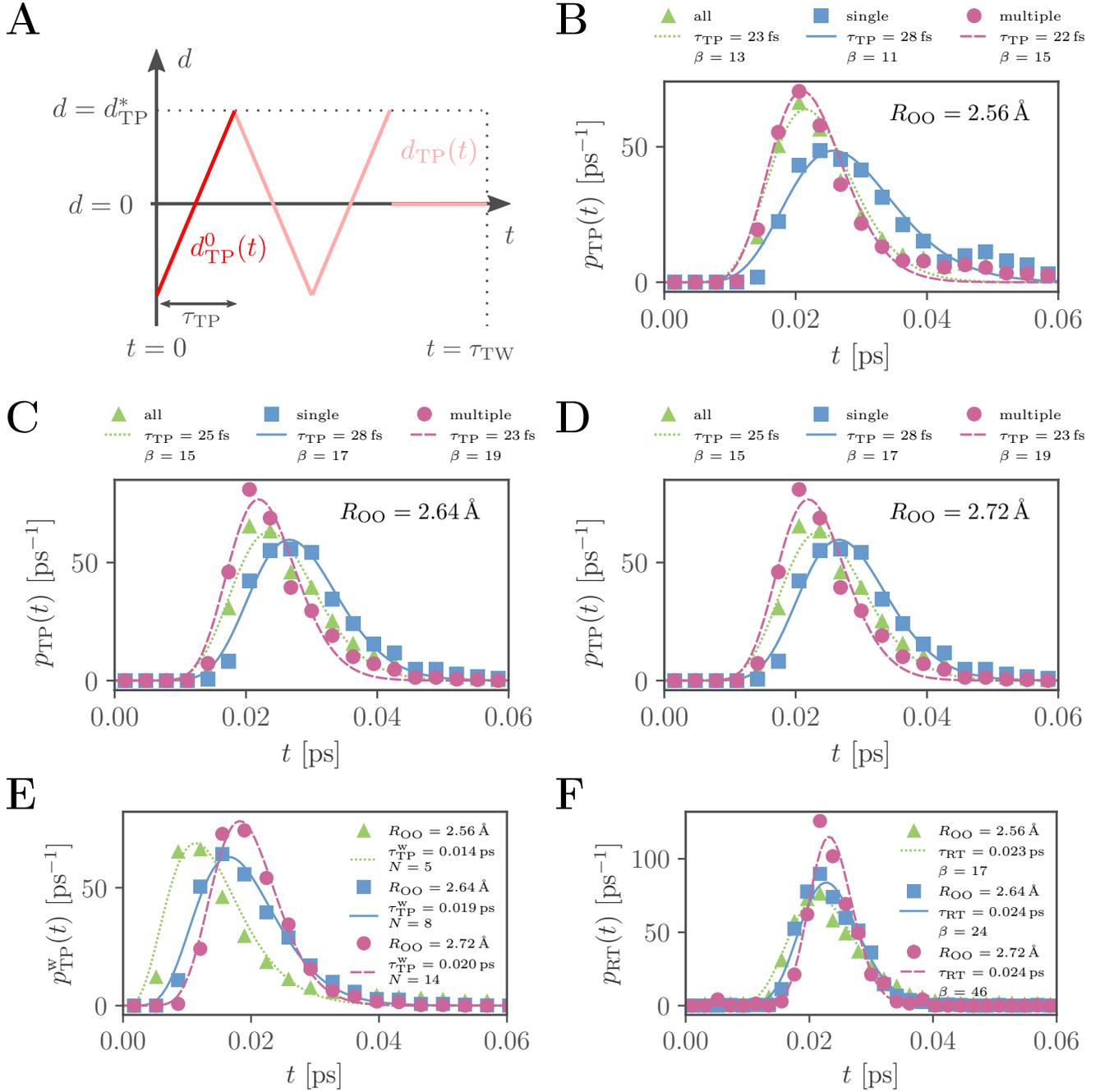

\begin{overpic}[width=0.49\textwidth]{{/../figs/tp_theory7}.pdf}
\put(-2,65){\huge \bf A}
\end{overpic}
\begin{overpic}[width=0.49\textwidth]{{/../figs3/tp/zundel_tp_d2.56_MinMaxTimes_ExpN}.eps}
\put(-2,65){\huge \bf B}
\end{overpic}
\begin{overpic}[width=0.49\textwidth]{{/../figs3/tp/zundel_tp_d2.64_MinMaxTimes_ExpN}.eps}
\put(-2,65){\huge \bf C}
\end{overpic}
\begin{overpic}[width=0.49\textwidth]{{/../figs3/tp/zundel_tp_d2.72_MinMaxTimes_ExpN}.eps}
\put(-2,65){\huge \bf D}
\end{overpic}
\begin{overpic}[width=0.49\textwidth]{{/../figs3/zundel_delta_fit_tps_short}.pdf}
\put(-2,57){\huge \bf E}
\end{overpic}
\begin{overpic}[width=0.49\textwidth]{{/../figs3/tp/zundel_tp_recrTimes_ExpN}.pdf}
\put(-2,57){\huge \bf F}
\end{overpic}
\caption{{\bf A} Model for the transfer-path trajectory $d_{\mathrm{TP}}(t)$ with length $\tau_{\mathrm{TW}}$. In this example the initial transfer path $d^0_{\mathrm{TP}}(t)$ with duration $\tau_{\mathrm{TP}}$ is centered around $t=0$ and followed by $N=2$ recrossing transfer paths, where $N$ is drawn from the recrossing-number probability distribution $p_{\mathrm{RN}}(N)$. The recrossing transfer path are alternating and shifted relative to each other by the recrossing time $\tau_{\mathrm{RT}} \approx \tau_{\mathrm{TP}}$.
{\bf B--D} Transfer-path-time probability distributions measured between the turning points ($p_{\mathrm{TP}}(t)$, B-D), between the minima of the free-energy landscape, ($p^{\mathrm{w}}_{\mathrm{TP}}(t)$, E) and recrossing-time distributions measured  between recrossing of $d=0$ ($p_{\mathrm{RT}}(t)$, F). The distributions (data points) are fitted to $p_{\mathrm{TP}}(t)=\frac{t^{\beta-1}}{(\beta-1)!} \left(\frac{\beta}{\tau}\right)^{\beta} e^{-\beta t/\tau}$ (solid, broken or dotted lines). Note that only the red curves in B--D are considered in the calculation of the TP spectral signature.}
\label{tpTheorySI}
\end{figure}

Recall that the power spectrum $\omega \widetilde \chi''(\omega)$ of a stochastic process $x(t)$ limited to the time domain $[0,L_t]$ is computed from the Fourier-transformed expressions $\tilde x(\omega)$ as presented in section \ref{linearResponseSection}. In the following an expression for $\tilde d_{\mathrm{TP}}(\omega)$ is derived, based on the  model  for $ d_{\mathrm{TP}}(t)$ illustrated in fig.~\ref{tpTheorySI}A. The mean transfer path is expected to repeat on average with period $\tau_{\mathrm{TW}}$. The model is therefore constrained to $[0,\tau_{TW}]$ and using eq.~\eqref{omegaDoublePrimeXOmega} the \ac{IR} power spectrum of the transfer-path contribution can then be written as
\begin{align}
\omega \widetilde \chi_{\mathrm{TP}}''(\omega)= \omega^2 \frac{q^2}{V\epsilon_0 k_BT \tau_{\mathrm{TW}}} \tilde d_{\mathrm{TP}}(\omega) \tilde d_{\mathrm{TP}}^*(\omega).
\label{chiTPSI}
\end{align}

$d_{\mathrm{TP}}(t)$ is modeled by a single mean transfer path with shape $ d^0_{\mathrm{TP}}(t)$ that is followed by a subsequent number of recrossing transfer paths, according to the recrossing-number probability distribution $p_{\mathrm{RN}}(n)$
\begin{align}
d_{\mathrm{TP}}(t) &=
\sum_{n=0}^{\infty} p_{\mathrm{RN}}(n) \sum_{m=0}^{n}\ (-1)^m d^{0}_{\mathrm{TP}}\left(t-m \tau_{\text{TP}} \right),
\label{xtTPtheoRecSI}
\end{align}
where the factor $(-1)^m$ accounts for the alternation of recrossing transfer paths that are going up and down. A Fourier transform with respect to $t$ turns the previous eq.~\eqref{xtTPtheoRecSI} into
\begin{align}
\tilde d_{\mathrm{TP}}(\omega) &= \sum_{n=0}^{\infty} p_{\mathrm{RN}}(n) \sum_{m=0}^{n}\ (-1)^m \int_{-\infty}^{\infty} dt\ e^{i \omega t}\ d^{0}_{\mathrm{TP}}(t-m\tau_{\text{TP}}) \nonumber\\
&= \tilde d^{0}_{\mathrm{TP}}(\omega)\ \sum_{n=0}^{\infty} p_{\mathrm{RN}}(n) \sum_{m=0}^{n}\ (-1)^m  e^{-i \omega m \tau_{\text{TP}}}.
\label{xwTPtheoRecSI}
\end{align}

\begin{figure}[tb]
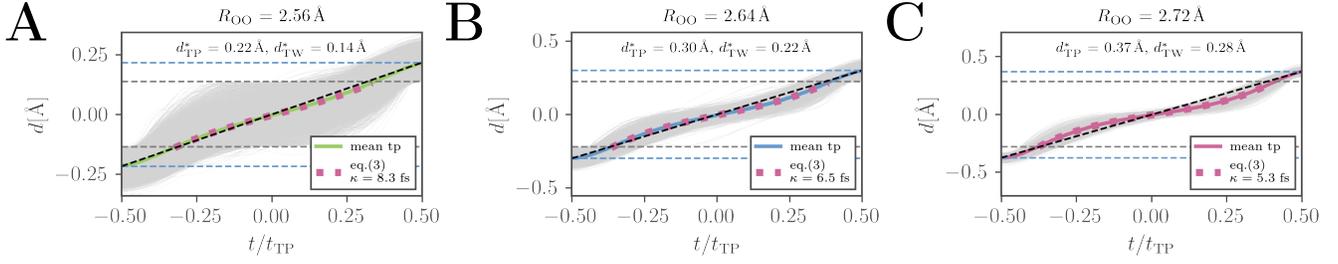

\begin{overpic}[width=0.32\textwidth]{{/../figs3/tp/zundel_tpFitMinMaxAvgD_d2.56}.png}
\put(-2,55){\huge \bf A}
\end{overpic}
\begin{overpic}[width=0.32\textwidth]{{/../figs3/tp/zundel_tpFitMinMaxAvgD_d2.64}.png}
\put(-2,55){\huge \bf B}
\end{overpic}
\begin{overpic}[width=0.32\textwidth]{{/../figs3/tp/zundel_tpFitMinMaxAvgD_d2.72}.png}
\put(-2,55){\huge \bf C}
\end{overpic}
\caption{Ensembles of transfer paths sampled from the trajectories of the excess proton in the H$_5$O$_2{}^{+}$ cation for different constrained oxygen distances $R_{\mathrm{OO}}$ ({\bf A}: $R_{\mathrm{OO}}=\SI{2.56}{\ang}$, {\bf B}: $R_{\mathrm{OO}}=\SI{2.64}{\ang}$, {\bf C}: $R_{\mathrm{OO}}=\SI{2.72}{\ang}$), scaled to their individual transfer-path times $t_{\mathrm{TP}}$ between the respective turning points. The blue dashed lines indicate the mean initial and final values, which are used to estimate the parameter $d^*_{\mathrm{TP}}$, the mean transfer-path distance along $d$. The horizontal black dashed lines indicate the minima of the free energy, which are used to estimate the parameter $d^*_{\mathrm{TW}}$. Mean transfer paths between the minima of the free energy (colored lines) are fitted using eq.~\eqref{eq:tpPathTheo}. The fits are shown as red dotted lines. Fit parameters are given in the legend.}
\label{allTransferPathsScaled}
\end{figure}

An expression for the mean transfer-path shape $ d^0_{\mathrm{TP}}(t)$ is obtained by regarding the ensembles of rescaled transfer paths, which are shown in fig.~\ref{allTransferPathsScaled} for various $R_{\mathrm{OO}}$, along with the mean transfer paths, obtained from space-averaging the ensembles of rescaled transfer paths at each rescaled time between  between the respective turning points of the trajectory. The mean value at the turning points defines the length scale $d^*_{\rm TP}$. The single mean transfer path $ d^0_{\mathrm{TP}}(t)$ reaching between the turning points is then modeled by a truncated straight line
\begin{align}
\label{xtTPtheoSingleLinSI}
d^0_{\mathrm{TP}}(t) &= 2 d^*_{\rm TP} \frac{t}{\tau_{\text{TP}}} \left(\theta \left(t\right)-\theta \left(t-\tau_{\text{TP}}\right)\right),
\end{align}
with the Fourier transform
\begin{align}
\label{xwTPtheoSingleLinSI}
\tilde d^0_{\mathrm{TP}}(\omega) &= d^*_{\rm TP} \frac{e^{i \omega  \tau _{\text{TP}}} \left(2-i \omega  \tau _{\text{TP}}\right)-2-i \omega  \tau _{\text{TP}}}{\omega ^2 \tau _{\text{TP}}}.
\end{align}

The time scale $\tau_{\text{TP}}$ in eq.~\eqref{xwTPtheoRecSI} is estimated directly from the simulation data, using the distributions, shown in fig.~\ref{tpTheorySI}B -- D. The distributions of transfer-path times considering only multiple transfers in fig.~\ref{tpTheorySI}B--D are sharply peaked at roughly the same value for all systems $\tau_{\mathrm{TP}}=\SI{0.023}{ps}$.

\begin{figure}[tb]
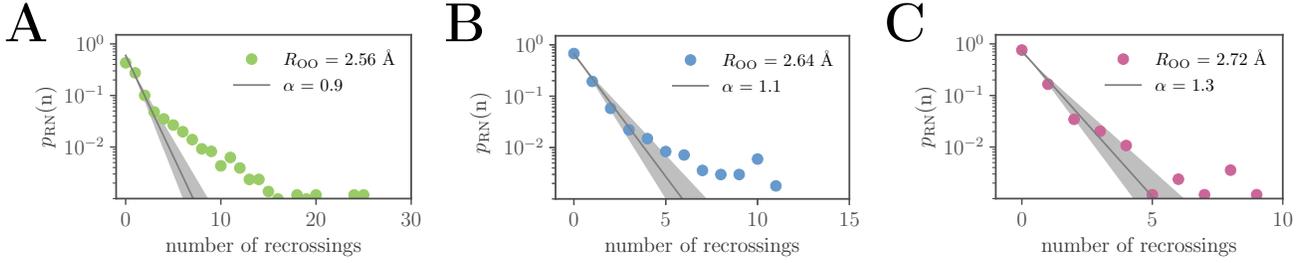

\begin{overpic}[width=0.32\textwidth]{{/../figs3/tp/zundel_tp_recrStat_expFit_d2.56}.pdf}
\put(-2,55){\huge \bf A}
\end{overpic}
\begin{overpic}[width=0.32\textwidth]{{/../figs3/tp/zundel_tp_recrStat_expFit_d2.64}.pdf}
\put(-2,55){\huge \bf B}
\end{overpic}
\begin{overpic}[width=0.32\textwidth]{{/../figs3/tp/zundel_tp_recrStat_expFit_d2.72}.pdf}
\put(-2,55){\huge \bf C}
\end{overpic}
\caption{Recrossing-number probability distributions $p_{\mathrm{RN}}(n)$ for different constrained oxygen distances $R_{\mathrm{OO}}$ ({\bf A}: $R_{\mathrm{OO}}=\SI{2.56}{\ang}$, {\bf B}: $R_{\mathrm{OO}}=\SI{2.64}{\ang}$, {\bf C}: $R_{\mathrm{OO}}=\SI{2.72}{\ang}$), normalized to $\sum_{n=0}^{\infty}p_{\mathrm{RN}}=1$ and fits according to $p_{\mathrm{RN}}(n)=(1-e^{-\alpha}) e^{- \alpha n}$. The grey shaded areas show the variation of $\alpha \pm 20\%$}
\label{recNumDist}
\end{figure}

To account for subsequent recrossings, the expression eq.~\eqref{xwTPtheoRecSI} is evaluated for an exponential decay of the recrossing-number probability distribution, $p_{\mathrm{RN}}(n) = (1-e^{-\alpha}) e^{- \alpha n}$, with decay parameter $\alpha$ and shown in fig.~\ref{recNumDist} for the given systems. Using this fit function the final expression for the transfer-path spectral contribution evaluates to
\begin{align}
\omega \widetilde \chi_{\mathrm{TP}}''(\omega)= \frac{{2 d^*_{\rm TP}}^2q^2}{V\epsilon_0 k_BT \tau_{\mathrm{TW}}}
\frac{e^{\alpha } \left(\omega  \tau _{\text{TP}} \cos \left(\frac{\omega  \tau _{\text{TP}}}{2}\right)-2 \sin \left(\frac{\omega  \tau _{\text{TP}}}{2}\right)\right){}^2}{\omega^2 \tau _{\text{TP}}^2 \left(\cosh (\alpha )+\cos \left(\omega  \tau _{\text{TP}}\right)\right)}.
\label{chiTPSILinG}
\end{align}

An approximate expression for eq.~\eqref{chiTPSILinG} is derived by factorizing $\omega \widetilde \chi_{\mathrm{TP}}''(\omega)$ into the recrossing contribution $X_{rec}(\omega, \alpha,\tau _{\text{TP}})$ and the shape contribution $|\tilde d^0_{\mathrm{TP}}(\omega,\tau_{\text{TP}})|^2$
\begin{align}
|\tilde d^{0}_{\mathrm{TP}}(\omega,\tau_{\text{TP}})|^2 &={d^*_{\rm TP}}^2 \frac{4 \left(\omega  \tau _{\text{TP}} \cos \left(\frac{\omega  \tau _{\text{TP}}}{2}\right)-2 \sin \left(\frac{\omega  \tau _{\text{TP}}}{2}\right)\right){}^2}{\omega ^4 \tau_{\text{TP}}^2},\\
X_{rec}(\omega, \alpha,\tau _{\text{TP}}) &= \frac{e^{\alpha }}{2 \left(\cosh (\alpha )+\cos \left(\omega  \tau _{\text{TP}}\right)\right)},
\label{chiTPRecG} \\
\omega \widetilde \chi_{\mathrm{TP}}''(\omega)&= \frac{q^2\omega^2}{V\epsilon_0 k_BT \tau_{\mathrm{TW}}} X_{rec}(\omega, \alpha,\tau _{\text{TP}}) |\tilde d^0_{\mathrm{TP}}(\omega,\tau_{\text{TP}})|^2.
\end{align}

The relevant maximum of $X_{rec}(\omega, \alpha,\tau _{\text{TP}})$ resides at $\omega = \pi / \tau_{\text{TP}}$, which minimizes the denominator in eq.\eqref{chiTPRecG}. A Taylor expansion of the $\cos$ function in eq.\eqref{chiTPRecG} to second order in $\omega$ around $\pi / \tau_{\text{TP}}$ leads to
\begin{align}
X_{rec}(\omega, \alpha,\tau _{\text{TP}}) &= \frac{e^{ \alpha }}{ 2 \cosh (\alpha ) - 2 +\left(\omega \tau _{\text{TP}} -\pi \right){}^2}.
\label{chiTPRecTayG}
\end{align}

Furthermore the shape contribution $|\tilde d^0_{\mathrm{TP}}(\omega,\tau_{\text{TP}})|^2$ can be estimated around $\omega=\pi / \tau_{\text{TP}}$ by the following expression, which is in good agreement with the local series expansion
\begin{align}
|\tilde d^{0}_{\mathrm{TP}}(\omega,\tau_{\text{TP}})|^2 &= {d^*_{\rm TP}}^2 \frac{64\omega ^2 \tau _{\text{TP}}^4}{\pi^4 \left(\omega  \tau _{\text{TP}}+\pi \right){}^2},
\label{chiTPSILinSTEG}
\\
\omega \widetilde \chi_{\mathrm{TP}}''(\omega)&= \frac{{d^*_{\rm TP}}^2q^2}{ V\epsilon_0 k_BT \tau_{\mathrm{TW}}}
\frac{64 e^{\alpha } \omega ^4 \tau _{\text{TP}}^4}{\pi^4 \left(\omega  \tau _{\text{TP}}+\pi \right){}^2 \left(2 \cosh (\alpha ) - 2 + \left(\omega \tau _{\text{TP}} -\pi \right){}^2 \right)}.
\label{chiTPSILinTEG}
\end{align}

Note that eq.~\eqref{chiTPSILinTEG} is a multiplication of a function of Debye-shape type with the shoulder frequency at $\omega=\pi/\tau{_\text{TP}}$, stemming from the shape contribution eq.~\eqref{chiTPSILinSTEG} and a function of Lorentz-shape type with the resonant frequency at $\omega=\pi/\tau{_\text{TP}}$, stemming from the recrossing contribution eq.~\eqref{chiTPRecTayG}.

\begin{figure*}
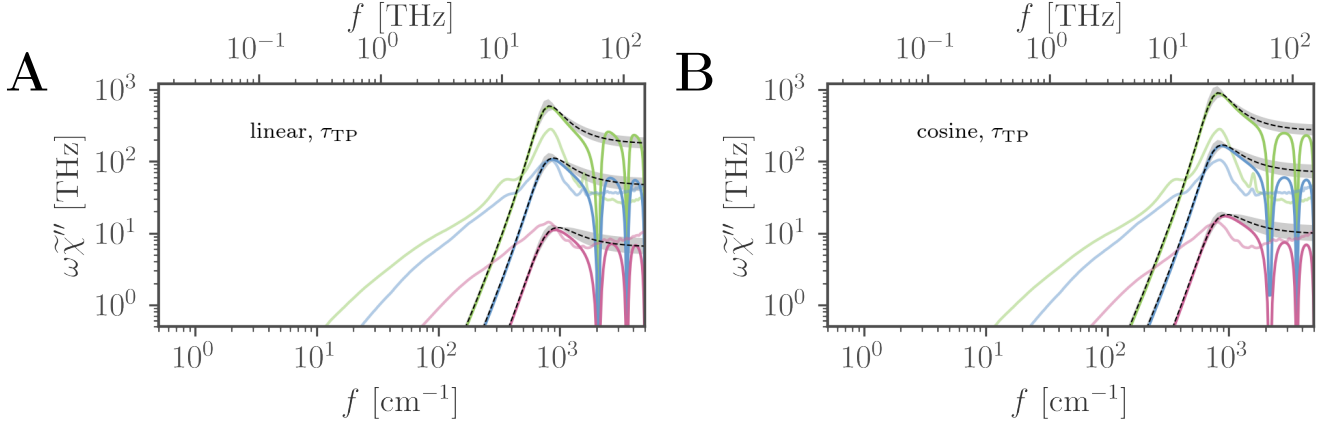

\begin{overpic}[width=0.49\textwidth]{{/../figs3/barrier/zundel_spectra_tpTheo_lin_all}.png}
\put(-2,55){\huge \bf A}
\end{overpic}
\begin{overpic}[width=0.49\textwidth]{{/../figs3/barrier/zundel_spectra_tpTheo_cos_all}.png}
\put(-2,55){\huge \bf B}
\end{overpic}
\caption{Models for the transfer-path spectral contribution given by eq.~\eqref{chiTPSILinG} for the linear transfer-path shape ({\bf A}) and eq.~\eqref{chiTPSICosG} for the cosine transfer-path shape ({\bf B}). The approximations are given by eqs.~\eqref{chiTPSILinTEG} and \eqref{chiTPSICosTEG} and are shown as black dashed lines in A and B. The grey shaded areas show the variation of $\alpha \pm 20\%$.}
\label{compareTPModels}
\end{figure*}

If the transfer-path shape $d^0_{\mathrm{TP}}(t)$ is alternatively modeled by a truncated cosine wave with period $2 \tau_{\text{TP}}$
\begin{align}
\label{xtTPtheoSingleSI}
d^0_{\mathrm{TP,cos}}(t) &= d^*_{\rm TP} \cos \left(\frac{\pi  t}{\tau_{\text{TP}}}\right) \left(\theta \left(t\right)-\theta \left(t-\tau_{\text{TP}}\right)\right),
\end{align}
with the Fourier transform
\begin{align}
\label{xwTPtheoSingleSI}
\tilde d^0_{\mathrm{TP,cos}}(\omega) &= -d^*_{\rm TP} \frac{i \omega  \tau _{\text{TP}}^2 \left(1+e^{i \omega  \tau _{\text{TP}}}\right)}{\omega ^2 \tau _{\text{TP}}^2-\pi ^2},
\end{align}
a slightly different result is obtained.

The final expression
\begin{align}
\omega \widetilde \chi_{\mathrm{TP,cos}}''(\omega)= \frac{{ d^*_{\rm TP}}^2q^2}{V\epsilon_0 k_BT \tau_{\mathrm{TW}}}
\frac{e^{\alpha } \omega ^4 \tau _{\text{TP}}^4 \left(\cos \left(\omega  \tau _{\text{TP}}\right)+1\right)}{\left(\pi ^2-\omega ^2 \tau _{\text{TP}}^2\right){}^2 \left(\cosh (\alpha
   )+\cos \left(\omega  \tau _{\text{TP}}\right)\right)},
\label{chiTPSICosG}
\end{align}
is Taylor-approximated as above to give
\begin{align}
\omega \widetilde \chi_{\mathrm{TP,cos}}''(\omega)&= \frac{{d^*_{\rm TP}}^2q^2}{ V\epsilon_0 k_BT \tau_{\mathrm{TW}}}
\frac{e^{\alpha } \omega ^4 \tau _{\text{TP}}^4}{\left(\omega  \tau _{\text{TP}}+\pi \right){}^2 \left(2 \cosh (\alpha ) - 2 + \left(\omega \tau _{\text{TP}} -\pi \right){}^2 \right)},
\label{chiTPSICosTEG}
\end{align}
which are both shown in fig.~\ref{compareTPModels}B. In fact both approximate equations differ only by a factor of $64/\pi^4 \approx 2/3$. The linear shape was chosen for the presentation in the main text because of the good qualitative agreement of the mean TP shape in fig.~\ref{allTransferPathsScaled} and quantitative agreement of the spectral shape in fig.~\ref{compareTPModels}A.

\newpage
\section{IR spectral decomposition of the excess-proton dynamics in the constrained H$_5$O$_2${}$^{+}$ cation}
\label{systemsSection}

\begin{figure*}[hb]
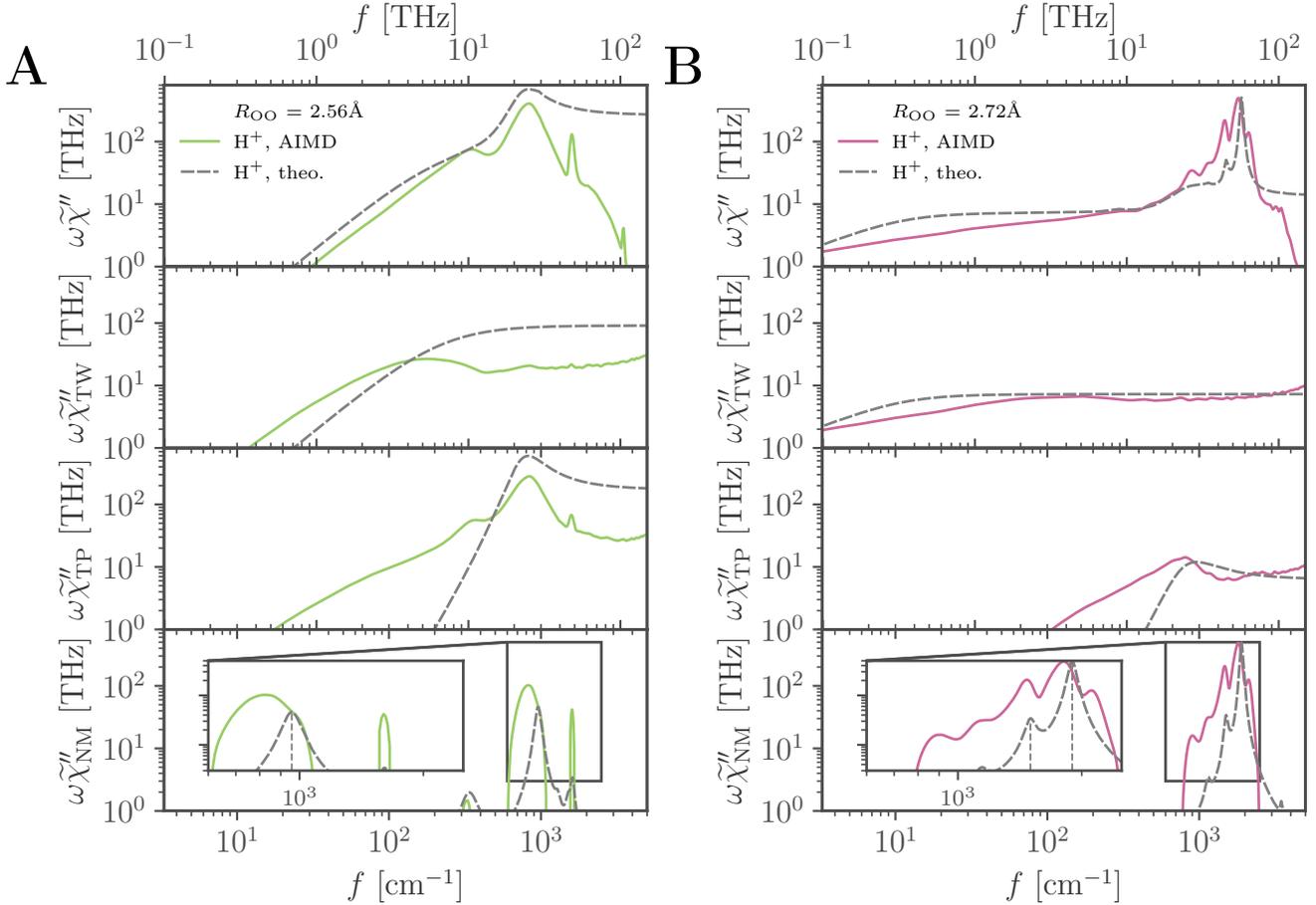

\centering
\begin{overpic}[width=0.49\textwidth]{{/../figs3/zundel_decompAllDtheo_d2.56_log}.pdf}
\put(-2,88){\huge \bf A}
\end{overpic}
\begin{overpic}[width=0.49\textwidth]{{/../figs3/zundel_decompAllDtheo_d2.72_log}.pdf}
\put(-2,88){\huge \bf B}
\end{overpic}
\caption{Spectral decomposition of the excess-proton dynamics in the H$_5$O$_2{}^{+}$ cation for different constrained $R_{\mathrm{OO}}$ ({\bf A}: $R_{\mathrm{OO}}=\SI{2.56}{\ang}$, {\bf B}: $R_{\mathrm{OO}}=\SI{2.72}{\ang}$) in analogy to the results in fig.~\ref{decomp}B in the main text (colored lines). Theoretical spectra for the models discussed in the previous sections are shown as grey broken lines.}
\label{systemsSI}
\end{figure*}

Spectral decompositions of the excess-proton dynamics in the H$_5$O$_2{}^{+}$ cation for different constrained $R_{\mathrm{OO}}$ are shown in fig.~\ref{systemsSI}. Theoretical spectra are shown for the barrier-crossing model derived in section \ref{binaryResponseSection} (respective eq.~\eqref{BarrierCrossingSpectrum} in the main text) and the transfer-path model derived in section \ref{theoryTPSection} (respective eq.~\eqref{TransferPathSpectrum} in the main text), using the distance between the minima of the free energy in fig.~\ref{systems}B in the main text and fits applied to the distributions in fig.~\ref{systems}C in the main text and in section \ref{theoryTPSection}.

\section{IR spectra of the deuterated H$_5$O$_2${}$^{+}$ cation}
\label{deuteronSpectraSection}

\begin{figure*}[hb]
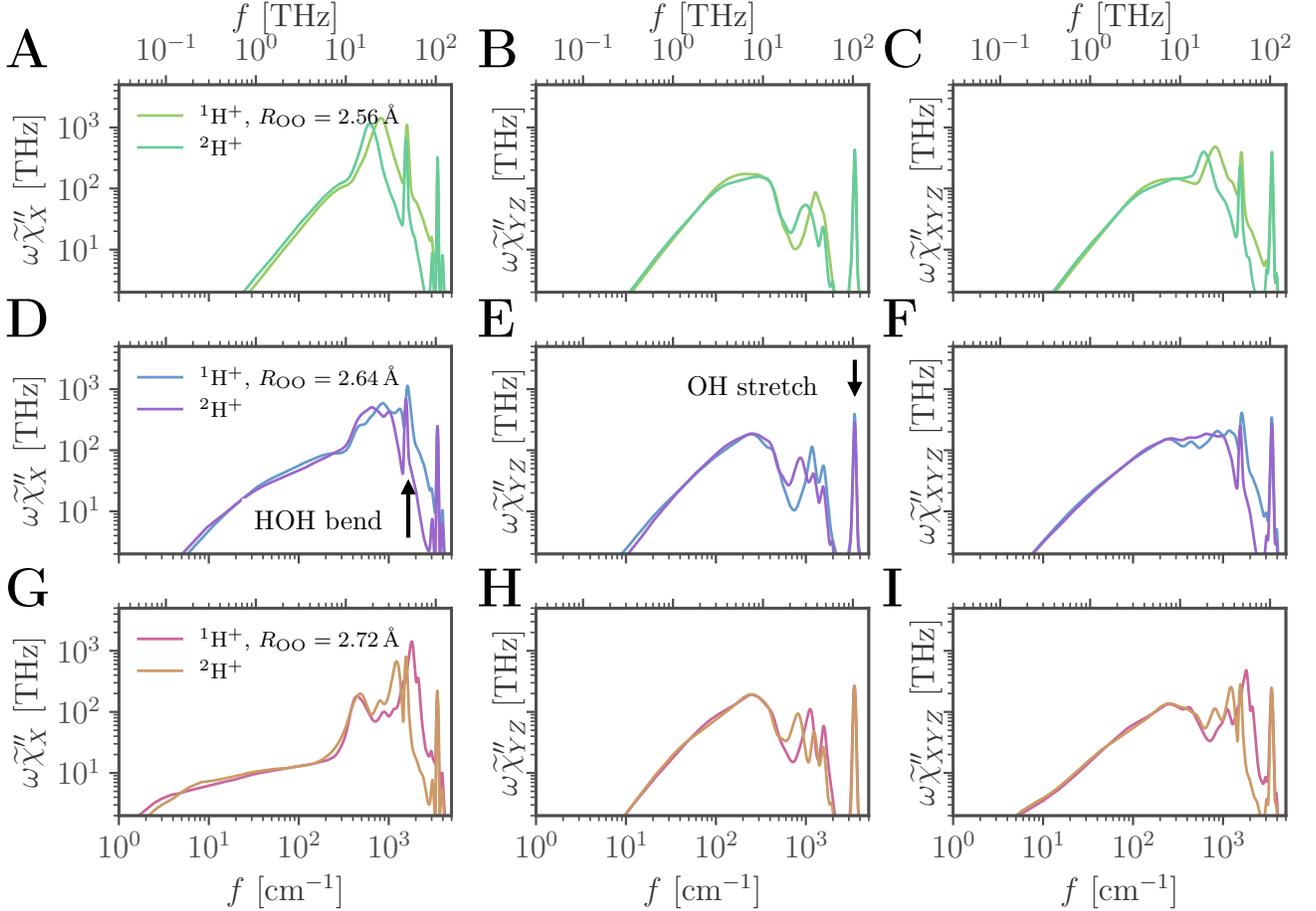

\centering
\begin{overpic}[width=\textwidth]{{/../figs/zundel_compare_spectra_d_xyz_all}.pdf}
\put(2,65){\huge \bf A}
\put(37,65){\huge \bf B}
\put(67,65){\huge \bf C}
\put(2,45){\huge \bf D}
\put(37,45){\huge \bf E}
\put(67,45){\huge \bf F}
\put(2,25){\huge \bf G}
\put(37,25){\huge \bf H}
\put(67,25){\huge \bf I}
\end{overpic}
\caption{Comparison of the \ac{IR} spectra of the H$_5$O$_2{}^{+}$ cation with an excess proton ($^1$H$^+$) or an excess deuteron ($^2$H$^+$) in different directions {\bf A, D, G}: along the $x$-axis connecting the two oxygens, {\bf B, E, H}: along the $yz$-plane {\bf C, F, I}: isotropic spectrum. Each row corresponds to a system with a distinct constrained oxygen distance $R_{\mathrm{OO}}$ given in the first legend. The HOH-bending mode and OH-stretching mode of the water molecules are indicated in {\bf D} and {\bf E}.}
\label{zundel_compare_spectra_d_xyz}
\end{figure*}

\begin{figure*}
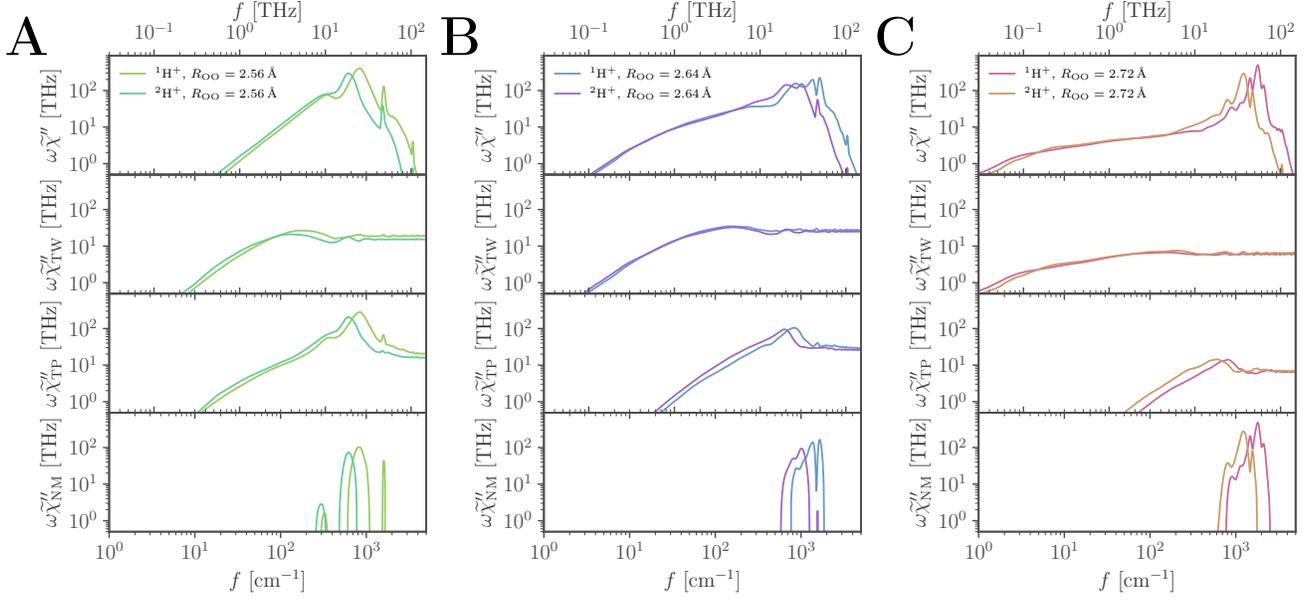

\centering
\begin{overpic}[width=0.32\textwidth]{{/../figs3/tp/zundel_tp_decompD_HD_d2.56_log}.pdf}
\put(-2,88){\huge \bf A}
\end{overpic}
\begin{overpic}[width=0.32\textwidth]{{/../figs3/tp/zundel_tp_decompD_HD_d2.64_log}.pdf}
\put(-2,88){\huge \bf B}
\end{overpic}
\begin{overpic}[width=0.32\textwidth]{{/../figs3/tp/zundel_tp_decompD_HD_d2.72_log}.pdf}
\put(-2,88){\huge \bf C}
\end{overpic}
\caption{Comparison of the \ac{IR} spectra of the H$_5$O$_2{}^{+}$ cation with an excess proton ($^1$H$^+$) or an excess deuteron ($^2$H$^+$) for different $R_{\mathrm{OO}}$ ({\bf A}: $R_{\mathrm{OO}}=\SI{2.56}{\ang}$, {\bf B}: $R_{\mathrm{OO}}=\SI{2.64}{\ang}$, {\bf C}: $R_{\mathrm{OO}}=\SI{2.72}{\ang}$), decomposed as described in the main text.}
\label{decompD}
\end{figure*}

In order to test isotope effects, simulations were performed with the excess proton replaced by an excess deuteron. The resulting \ac{IR} spectra for various $R_{\mathrm{OO}}$ and directions are shown in fig.~\ref{zundel_compare_spectra_d_xyz} and compared to the excess-proton data. The regime between \SIrange{400}{2000}{cm^{-1}} shows various shifts to lower frequencies and also splittings. This H/D isotope effect is expected since this region is linked to the excess-proton motion. As expected, the OH-stretching vibrations are not effected. These vibrations are associated with the flanking water molecules, that are not deuterated. Fig.~\ref{decompD} compares the H/D isotope effect for the decomposed spectra. In contrast to the normal-modes involving the excess-proton as well as the transfer-path signature, which are all shifted by deuteration, the low-frequency transfer-waiting shoulder does not shift.

\section{Deuteron transfer-waiting rates}
\label{deuteronRatesSection}

\begin{figure*}
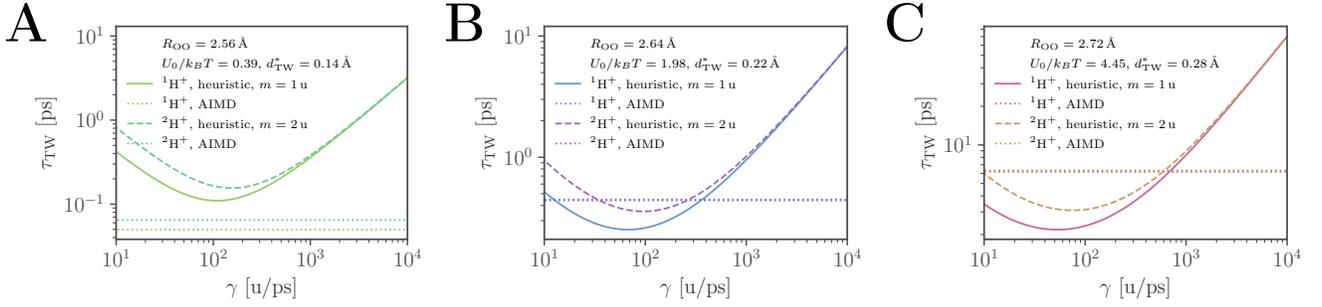

\centering
\begin{overpic}[width=0.32\textwidth]{{/../figs3/d/rates/kramers_d2.56}.pdf}
\put(-2,65){\huge \bf A}
\end{overpic}
\begin{overpic}[width=0.32\textwidth]{{/../figs3/d/rates/kramers_d2.64}.pdf}
\put(-2,65){\huge \bf B}
\end{overpic}
\begin{overpic}[width=0.32\textwidth]{{/../figs3/d/rates/kramers_d2.72}.pdf}
\put(-2,65){\huge \bf C}
\end{overpic}
\caption{The transfer-waiting times $\tau_{\mathrm{TW}}$ of the excess proton ($^1$H$^+$) and the excess deuteron ($^2$H$^+$) of the H$_5$O$_2{}^{+}$ cation for different constrained oxygen distances $R_{\mathrm{OO}}$ (dotted lines, {\bf A}: $R_{\mathrm{OO}}=\SI{2.56}{\ang}$, {\bf B}: $R_{\mathrm{OO}}=\SI{2.64}{\ang}$, {\bf C}: $R_{\mathrm{OO}}=\SI{2.72}{\ang}$) compared to the heuristic formula, given by eq.~\eqref{KapplerFormula}, (straight and broken lines) as a function of the friction constant $\gamma$.  Parameters entering the heuristic formula are given in the legends.}
\label{zundel_compare_rates}
\end{figure*}

As discussed in the previous section, no isotope effect is found for the low-frequency transfer-waiting shoulder of the \ac{IR} spectrum, i.e. the mean transfer-waiting time $\tau_{\mathrm{TW}}$ does not depend on the mass of the reaction coordinate, but solely on friction constant $\gamma$. This is expected to be the same for the system with either an excess proton or an excess deuteron. The determination of the friction constant $\gamma$ is not straight-forward. In fig.~\ref{zundel_compare_rates} the mean transfer-waiting times $\tau_{\mathrm{TW}}$ measured from the simulations are compared to the heuristic formula \cite{Kappler2019}
\begin{align}
\tau_{\mathrm{TW}} =  e^{\frac{U_0}{k_BT}} \left[ \frac{m}{\gamma} \frac{k_BT}{U_0} + \frac{\pi}{2\sqrt{2}}\frac{\gamma {d^*_{\rm TW}}^2}{U_0}+2\sqrt{\frac{m{d^*_{\rm TW}}^2}{U_0}} \right].
\label{KapplerFormula}
\end{align}
Since the mass $m$, barrier height $U_0$ and widths of the barrier $2{d^*_{\rm TW}}$ are known (given in fig.~\ref{zundel_compare_rates}), estimates for the friction constant $\gamma$ are obtained by comparing the simulation data to the heuristic formula. The obtained values, $\gamma \approx \SI{400}{u/ps}$ for $R_{\mathrm{OO}}=\SI{2.64}{\ang}$ and $\gamma \approx \SI{900}{u/ps}$ for $R_{\mathrm{OO}}=\SI{2.72}{\ang}$, vary greatly and also deviate from the friction constants fitted to  the line-broadening of the normal-modes, $\gamma = \SI{16}{u/ps}$. This discrepancy highlights the complex frequency dependence of dielectric friction, as previously discussed by \citet{Sedlmeier2014, Brunig2022}. Note that the theoretical predictions are only valid for $U_0 > 2\,k_BT$, and can therefore not be used for interpretation of the data for the lowest barrier considered here \cite{Kappler2019}.
Nevertheless, fig.~\ref{zundel_compare_rates} validates the observation of the negligible isotope effect on the mean transfer-waiting time, since the heuristic formula of eq.~\eqref{KapplerFormula} shows little mass dependence in the predicted regime of $\gamma$, as seen in figs.~\ref{zundel_compare_rates}B and \ref{zundel_compare_rates}C from the negligible difference between the curves for $m=\SI{1}{u}$ (solid lines) and $m=\SI{2}{u}$ (dashed lines) in the regime of the measured mean transfer-waiting time (dotted lines).

\clearpage

\section{Wiener-Kintchine theorem}
\label{WienerKintchineSection}

The correlation function $C_{xy}(t)$ of two stochastic processes $x(t)$ and $y(t)$ limited to the interval $[0,L_t]$ is efficiently computed from the Fourier-transformed expressions $\tilde x(\omega)$ and $\tilde y(\omega)$ according to
\begin{align}
\label{WienerKintchine}
C_{xy}(t) = \frac{1}{2 \pi (L_t-t)} \int_{-\infty}^{\infty} d\omega\ e^{-i\omega t} \tilde x(\omega) \tilde y^*(\omega),
\end{align}
where the asterix denotes the conjugate form. This is known as the Wiener-Kintchine theoreme \cite{Wiener1930}. Both sides of eq.~\eqref{WienerKintchine} are Fourier-transformed to give
\begin{align}
\int_{-\infty}^{\infty} dt\ e^{i\omega t}\ L_t \left( 1-\frac{t}{L_t}\right) C_{xy}(t) &= \tilde x(\omega) \tilde y^*(\omega),
\end{align}
which in the limit of large $L_t$ reduces to
\begin{align}
\label{WienerKintchineFT}
\widetilde C_{xy}(\omega) &= L_t^{-1} \tilde x(\omega) \tilde y^*(\omega).
\end{align}
Eq.~\eqref{WienerKintchine} can be derived starting off with the definition of the correlation function
\begin{align}
C_{xy}(t) = \frac{1}{L_t-t}\int_{0}^{L_t-t} dt'\ x(t'+t) y(t').
\end{align}
Defining $x(t), y(t) = 0$ for $t \not\in [0, L_t ]$, the integral bounds can formally be extended
\begin{align}
  C_{xy}(t) = \frac{1}{L_t-t}\int_{-\infty}^{\infty} dt'\ x(t'+t) y(t'),
  \end{align}
and making use of the convolution theorem
\begin{align}
C_{xy}(t) &= \frac{1}{4 \pi^2 (L_t-t)} \int_{-\infty}^{\infty} dt' \int_{-\infty}^{\infty} d\omega\ e^{-i\omega (t+t')} \tilde x(\omega) \int_{-\infty}^{\infty} d\omega'\ e^{-i\omega' t'} \tilde y(\omega') \nonumber \\
 &= \frac{1}{4 \pi^2 (L_t-t)} \int_{-\infty}^{\infty} d\omega\ e^{-i\omega t} \tilde x(\omega) \int_{-\infty}^{\infty} d\omega'\ \tilde y(\omega')
 \int_{-\infty}^{\infty}  dt'  e^{-it' (\omega+\omega')} \nonumber \\
 &= \frac{1}{4 \pi^2 (L_t-t)} \int_{-\infty}^{\infty} d\omega\ e^{-i\omega t} \tilde x(\omega) \int_{-\infty}^{\infty} d\omega'\ \tilde y(\omega')
2 \pi \delta(\omega+\omega') \nonumber \\
 &= \frac{1}{2 \pi (L_t-t)} \int_{-\infty}^{\infty} d\omega\ e^{-i\omega t} \tilde x(\omega) \tilde y(-\omega),
\end{align}
noting that $\tilde y(-\omega)=\tilde y^*(\omega)$ for a real function $y(t)$ in order to obtain eq.~\eqref{WienerKintchine}.

\begin{acronym}[Bash]
  \acro{AIMD}{ab initio molecular-dynamics}
  \acro{DFT}{density functional theory}
  \acro{FPT}{first-passage time}
  \acro{GLE}{generalized Langevin equation}
  \acro{GH}{Grote-Hynes}
  \acro{IR}{infrared}
  \acro{LE}{Langevin equation}
  \acro{MD}{molecular dynamics}
  \acro{MFPT}{mean first-passage time}
  \acro{MFP}{mean first-passage}
  \acro{MSD}{mean squared displacement}
  \acro{NM}{normal-mode}
  \acro{PTP}{$p(\text{TP}|q)$}
  \acro{PME}{particle mesh Ewald~\cite{pronk2013gromacs}}
  \acro{PMF}{potential of mean force}
  \acro{PGH}{Pollak-Grabert-Hanggi}
  \acro{RC}{reaction coordinate}
  \acro{RDF}{radial distribution function}
  \acro{RTT}{round-trip time}
  \acro{TP}{transfer path}
 \end{acronym}
 
 % Create the reference section using BibTeX:
\bibliography{bibliography_si.bib}